\documentclass[
    reprint,
    amsmath,
    amssymb,
    aps,
    floatfix,
    longbibliography
]{revtex4-2}

\usepackage{graphicx} 
\usepackage{hyperref}
\usepackage{prettyref}
\usepackage{amsmath,amssymb}
\usepackage{bm}       
\usepackage{xcolor}
\usepackage{tikz}
\usepackage{upgreek}
\usepackage{booktabs,array,multirow}
\usepackage{placeins}

\newrefformat{eq}{\hyperref[#1]{Eq.~(\ref*{#1})}}
\newrefformat{eqs}{\hyperref[#1]{Eqs.~(\ref*{#1})}}
\newrefformat{fig}{\hyperref[{#1}]{Figure~\ref*{#1}}}

\newrefformat{tab}{\hyperref[#1]{Table ~\ref*{#1}}}
\newrefformat{chap}{\hyperref[#1]{Chapter~\ref*{#1}}}
\newrefformat{sec}{\hyperref[#1]{Section~\ref*{#1}}}
\newrefformat{subsec}{\hyperref[#1]{Subsection~\ref*{#1}}}
\newrefformat{app}{\hyperref[#1]{Appendix~\ref*{#1}}}
\newrefformat{par}{\hyperref[#1]{§~\ref*{#1}}}
\newrefformat{app}{\hyperref[#1]{Appendix~\ref*{#1}}}

\newcommand{\cref}[1]{{\color{blue}\prettyref{#1}}}
\newcommand{\ccite}[1]{{\color{blue}\cite{#1}}}

\newcommand{\sm}[1]{ \text{\small{${#1}$}}}

\makeatletter
\newcommand{\ostar}{\mathbin{\mathpalette\make@circled *}}
\newcommand{\make@circled}[2]{%
  \ooalign{$\m@th#1\smallbigcirc{#1}$\cr\hidewidth$\m@th#1#2$\hidewidth\cr}%
}
\newcommand{\smallbigcirc}[1]{%
  \vcenter{\hbox{\scalebox{0.77778}{$\m@th#1\bigcirc$}}}%
}

\definecolor{ForestGreen}{RGB}{34,139,34}
\definecolor{blue-violet}{rgb}{0.54, 0.17, 0.89}
\definecolor{blue(pigment)}{rgb}{0.2, 0.2, 0.6}
\definecolor{amber(sae/ece)}{rgb}{1.0, 0.49, 0.0}
\definecolor{amber}{rgb}{1.0, 0.75, 0.0}
\definecolor{antiquefuchsia}{rgb}{0.57, 0.36, 0.51}
\definecolor{aurometalsaurus}{rgb}{0.43, 0.5, 0.5}
\definecolor{arsenic}{rgb}{0.23, 0.27, 0.29}
\definecolor{ballblue}{rgb}{0.13, 0.67, 0.8}
\definecolor{darkblue}{rgb}{0.0, 0.0, 0.3}
\definecolor{goldenpoppy}{rgb}{0.99, 0.76, 0.0}
\definecolor{orangered}{rgb}{1,0.271, 0}
\definecolor{amber(sae/ece)}{rgb}{1.0, 0.49, 0.0}

\definecolor{processesAll_Solv}{HTML}{0072BD}
\definecolor{processesAll_Ic}{HTML}{7E2F8E}
\definecolor{processesAll_Ssp}{HTML}{D95319}
\definecolor{processesAll_Prec}{HTML}{EDB120}
\definecolor{processesAll_Pvk}{HTML}{A2142F}

\definecolor{evaporation}{RGB}{27,187,255}
\definecolor{chemicalReactions}{RGB}{112,48,160}
\definecolor{crystallization}{RGB}{217,103,9}

\newcommand{\labelpanel}[3][scale=0.22]{%
  \begin{tikzpicture}
    \node[anchor=south west, inner sep=0] (img) at (0,0) {\includegraphics[#1]{#2}};
    \node[anchor=north west, font=\bfseries, inner sep=0pt] at ([xshift=0pt,yshift=0pt]img.north west) {#3)};
  \end{tikzpicture}%
}

\newcommand{\sweep}{\ensuremath{\textit{sw}}\,}   

\begin{document}

\title{Coupling Evaporation, Chemical Reactions, and Crystallization: A Population-Balance Framework Applied to Two-Step Perovskite Crystallization }
\author{Kai Segadlo}
\email{k.segadlo@fz-juelich.de}
\affiliation{Helmholtz Institute Erlangen-N\"urnberg for Renewable Energy (IET-2), Forschungszentrum J\"ulich, F\"urther Straße 248, 90429 N\"urnberg, Germany}
\author{Olivier J.J. Ronsin}
\affiliation{Helmholtz Institute Erlangen-N\"urnberg for Renewable Energy (IET-2), Forschungszentrum J\"ulich, F\"urther Straße 248, 90429 N\"urnberg, Germany}
\author{Jens Harting}
\email{j.harting@fz-juelich.de}
\affiliation{Helmholtz Institute Erlangen-N\"urnberg for Renewable Energy (IET-2), Forschungszentrum J\"ulich, F\"urther Straße 248, 90429 N\"urnberg, Germany}
\affiliation{Department of Chemical and Biological Engineering and Department of Physics, Friedrich-Alexander-Universität Erlangen-Nürnberg, Fürther Straße 248, 90429 Nürnberg, Germany}
\date{\today}

\begin{abstract}
A limited understanding of how evaporation couples to crystallization pathways hampers control over the formation of solution-processed perovskite photoactive layers. Identifying these mechanisms—and learning how to select and tune them—is essential for achieving the desired crystal morphology and, ultimately, improving solar-cell performance and stability. To investigate crystallization pathways in evaporating thin films, we develop a simple and general zero-dimensional population-balance model that couples evaporation, chemical reactions, and crystallization. We use the model to explore a range of parameters for a solvent–solute system that follows a two-step crystallization pathway: solvent evaporation first induces formation of a crystalline precursor, which subsequently recrystallizes into the final phase. The model predicts the evolution of the film composition during drying and provides quantitative LaMer curves. Based on these results, we introduce a vocabulary of “supply limitations” and “rate limitations” that enables the interpretation of crystallization kinetics in terms of the governing process conditions.
\end{abstract}
\maketitle

\section{Introduction}
Solar cells are a primary energy-harvesting technology, with steadily decreasing energy costs per kilowatt-hour and, in many regions, a lower levelized cost of electricity than fossil-based generation. They contribute significantly to the renewable energy transition~\ccite{RenewablePowerGeneration2022}.
Solar cells with the potential for solution processing that avoid costly traditional high-vacuum, high-temperature fabrication conditions are progressing toward pilot-scale and early-stage industrial production~\ccite{almoraDevicePerformanceEmerging2025}. Among the solution-processable photoactive layer materials, for which roll-to-roll production-compatible techniques, such as slot-die coating, have been demonstrated, metal halide perovskites are a promising class. This owes to their compositionally tunable optoelectronic properties and the high power conversion efficiencies of the resulting solar cells~\ccite{leeCuttingedgeAdvancesScalable2024}.
However, solution processing of thin films poses fundamental challenges for controlling their formation. For perovskites, the films' optoelectronic properties are closely linked to film roughness, grain size, density, and grain-boundary density. 
Thus, achieving reproducible, durable, high-performing devices requires precise control of crystallization~\ccite{yangUpscalingSolutionProcessedPerovskite2021,Qiu-EES-2025}. 
This control,  however, is challenging; in many deposition protocols, solvent evaporation dynamically controls crystallization, triggering nucleation after a critical solution concentration is reached, and determines the time evolution of the concentration-dependent nucleation and crystal growth rates. The chemical activity of the precursor solutions further amplifies this complexity.

To elucidate film formation mechanisms, a range of in situ characterization techniques has been used. Early studies relied primarily on ex situ X-ray diffraction (XRD) to analyze the final crystalline phases and infer crystallization pathways retrospectively~\ccite{WangGrowthDegradationRew2021}. More recently, time-resolved in situ techniques—including grazing-incidence wide-angle X-ray scattering (GIWAXS)~\ccite{steeleHowGIWAXSGrazing}, optical laser or white light reflectometry (LR/WLR)~\ccite{schmidt-hansberg_spatially_2011, rappich_fast_2021-1}, and combined multimodal approaches~\ccite{dasilvavanessaMultimodalDataAcquisition2025}—have enabled direct observation of intermediate phases and phase transitions during drying and annealing~\ccite{szostakSituOperandoCharacterizations2023}. These studies have significantly advanced the mechanistic understanding of perovskite crystallization~\ccite{ternes_drying_2019-2, merdasa_eye_2020, alhazmi_perovskite_2020, shargaievaTemperatureDependentCrystallizationMechanisms2021, pratapOutofequilibriumProcessesCrystallization2021, rehermann_role_2022, reusTimeResolvedOrientationPhase2022, ternesDryingCoatingPerovskite2022}. Strong solvent–precursor interactions can lead to the formation of colloidal pre-clusters~\ccite{pratapHybridPerovskitesColloidal2021} and solvent-coordinated crystalline intermediate phases prior to the final perovskite. Intermediates alter supersaturation dynamics and provide alternative nucleation pathways: The perovskite may form directly, in parallel to, or from intermediate phases~\ccite{schmidt-mendeRoadmapOrganicInorganic2021}. Consequently, the final film morphology—including grain size distribution, grain boundary density, and defect density—is highly sensitive to the specific crystallization pathway followed during deposition~\ccite{hwang_challenges_2023,ortoll-blochBypassingSolidStateIntermediates2020}. Despite these advances in in-situ monitoring, many investigations remain incomplete and qualitative~\ccite{ternesProcessParameterSpecification}. In particular, evaporation kinetics, which control crystallization, are not consistently measured. Moreover, suggested film-formation pathways are rarely quantitatively tested against the data. 

In parallel, modeling frameworks that integrate chemical reactions, solvent evaporation, and crystallization are lacking in photovoltaic research. Quantitative modeling is rare and typically relies on decoupled, substantially simplified process descriptions. Evaporation of two solvents from a perovskite precursor solution has been successfully modeled under realistic coating conditions in~\ccite{ternes_drying_2019-2,ternesDryingCoatingPerovskite2022,ternesSituCharacterizationModelling2022} but without accounting for the solidification of the film. Furthermore, a single secondary perovskite crystallization could be modeled as a single process in~\ccite{suchanOpticalSituMonitoring2020} and~\ccite{mrkyvkovaCombinedSituPhotoluminescence2021}, but the role of evaporation and prior solid-state precursor crystallization has not been likewise considered. In other fields, for instance, chemical engineering, more holistic approaches are employed. For chemical reactions, for example, automated model discrimination provides systematic strategies for assessing reaction pathways using experimental data~\ccite{taylorRapidAutomatedDetermination2021}. For particulate systems, population balance modeling (PBM) can serve as a framework to describe how multiple distinct processes, such as chemical reactions, evaporation, and crystallization, jointly affect particle populations (e.g., crystals)~\ccite{B0_rawlingsModelIdentificationControl1993}. For instance, it has been used for direct crystallization control of a polymer~\ccite{verkiCrystallizationParticleSize2025}, to understand salt crystallization in droplets~\ccite{hardyAccurateMeasurementsSimulations2023}, or to rationalize the growth of single perovskite crystals at constant supersaturation~\ccite{gaoRoomTemperatureSolventEvaporation2023}. 
However, such integrated approaches have not yet been widely applied to the formation of perovskite thin films. Applied modeling is rare and typically relies on decoupled or overly simplified process descriptions~\ccite{poolThermalEngineeringFAPbI32017,suchanOpticalSituMonitoring2020,mrkyvkovaCombinedSituPhotoluminescence2021}.
A general population balance model framework to clarify the role of evaporation in crystallization and perovskite crystallization pathways is still lacking.

Therefore, given the experimental findings on hypothetical mechanisms and the modeling state outlined above, we conclude that it is useful to propose a model that can simultaneously handle multiple chemical reactions, the evaporation of multiple solvents, and multiple successive or competitive crystallization processes. 

In this paper, we propose a general zero-dimensional kinetic model for the evolution of the material composition in reactive, evaporating, and crystallizing thin films. The model, in principle, allows us to quantitatively compare a variety of process systems encompassing multiple coupled evaporation, crystallization, and chemical processes. Subsequently, in a simulation study, we investigate evaporation-induced successive crystallization for perovskite formation.

\section{Model description}
In this section, we detail our general modeling framework. It allows us to describe a set of processes (chemical reactions, evaporation, crystallization, sublimation, condensation, light-, or stress-induced evolution, demixing,...), over a set of states defined, for instance, by phase (e.g., gaseous $(g)$, liquid $(\ell)$, or solid $(s)$) and molecular species (e.g. $\text{H}_2\text{O}$).

First, we outline the theory's general idea. Then, for each process type, we go through the basic building blocks and the rate expressions in turn.
\subsection{Population balance model} 
Consider a process $p$ converting $n$ reagents $A_i$ into $m$ distinct products $A_j$ with all $i \neq j$:
\begin{align*}
    a_1 \, A_1 + \dots a_n A_n \to   a_{n+1} \, A_{n+1} + \dots + a_{n+m} \, A_{n+m} 
\end{align*}
Here, $a_1,\dots, a_{n+m}$ are the reactant and product stochiometric coefficients. In the following, we consider only chemical reactions, crystallization, and evaporation processes, but these can, in principle, be replaced by any other process (sublimation, condensation, light- or stress-induced evolution, demixing, ...). Thereby, in this work, the reagents and products are defined by their chemical nature and phase state (e.g., chemical species in the liquid phase, a gaseous solvent, atoms or molecules in the crystalline state, etc.). However, more generally, they can also be defined by additional properties, such as mass or size (e.g., a polymer chain of given molar mass, a particle of given volume, ...).
Then, the population balance for the number of moles $n$ of all reagents and products is given by the process advancement rate $\xi_p$ of process $p$:  
\begin{align*}
     \xi_p &= - \frac{1}{a_1 }\frac{dn_{1}}{dt} =  \dots =- \frac{1}{a_n }\frac{dn_{n}}{dt} \\
     &=  \frac{1}{a_{n+1}}\frac{dn_{n+1}}{dt} = \dots =   \frac{1}{a_{n+m}}\frac{dn_{n+m}}{dt}
\end{align*}

This allows us to obtain the contribution of process $p$ to the population evolution of any reagent or product $k$ as
\begin{align*}
    \frac{dn_{k,p}}{dt} &=  \epsilon_p a_k \,  \xi_p ,  
\end{align*}
whereby the sign $\epsilon_p$ is negative or positive depending on whether $k=1...(n+m)$ appears as a reactant or product.

Finally, if there are $N_p$ active processes in the system, the global population evolution of each species is determined by the evolution of all processes. More specifically, if a species $k$ is involved as either reactant or product in $N_{p,k}$ of the $N_p$ processes with stochiometry $\epsilon_p a_{k,p}$, its population changes as
\begin{align*}
    \frac{dn_{k}}{dt} =  \sum\limits_{p=1}^{N_{p,k}} \frac{dn_{k,p}}{dt} =  \sum\limits_{p=1}^{N_{p,k}} \epsilon_p a_{k,p}  \xi_p .
\end{align*} 

The equation above expresses the population balance for all species subject to the set of active processes and enables evaluation of the time-dependent evolution of species populations.

In the following, we detail the process-specific rates $\xi_p$ for chemical reactions, evaporation, and crystallization. Thereby, the starting point for a process $p$ will typically be an empirical volumetric process advancement rate $\frac{\xi_p}{V_{M,p}}$. This rate is measured with reference to the volume $V_{M,p}$ within which the process takes place, which we refer to as the "mother phase volume" hereafter. The definition of the mother phase depends on the process and will be detailed below.

\subsection{Chemical reactions}
As a kinetic model for chemical reactions, we use a generalization of the of mass-action rate
equations~\ccite{bazant2013theory} formulated through chemical potentials rather than concentrations~\ccite{vallanceIntroductionChemicalKinetics2017}. 
For a chemical reaction involving $m$ reactants $A_j$, with stochiometry $a_j$ with $j=1...m$, whose chemical potentials
$\mu_j^{(\ell)}$ in the liquid phase depend on the volume fractions $\phi_j$, the chemical reaction rate $\xi_\chi$ is
\begin{align*}
     \xi_\chi = V_M \,\kappa  \prod_{j=1}^m e^{ a_j \mu_j^{(\ell)} /RT}.
\end{align*}
For our target application, we  consider reactions in a common liquid phase $V_M=V^{(\ell)}$. Here, $\kappa$ is an Arrhenius-type rate prefactor, assumed independent of composition, and the reaction orders are taken to equal the stoichiometric coefficients $a_j$.

To characterize the mixture thermodynamics, we assume an incompressible liquid that can be described by the Flory-Huggins-Gibbs free energy density
\begin{align*}
    f_{V}^{(\ell)} &= \tfrac{R T}{\nu_0} \left( \sum_i \tfrac{1}{N_i}\phi_i\ln \phi_i + \tfrac{1}{2} \sum_{ij}  \phi_i \chi_{ij} \phi_j \right) ,
\end{align*}  
where the $\chi_{ij} =\chi_{ji}, \chi_{ii} =0$ denote pairwise material volume interaction parameters and $N_i = \nu_i/ \nu_0$ relative molar volumes with molar volumes $\nu_i$ and reference $\nu_0 = \min_j{\nu_j}$. From the free energy density, we deduce the material-resolved chemical potentials for the liquid part of the blend: 
\begin{align*}
    \mu_i^{(\ell)}/RT
        &=  \Big(&&\phantom{+}1 + \ln \phi_i   -\sm{N_i}  \sum_j \tfrac{1}{N_j}\,\phi_j &&&\\
    &  &&+  \sm{N_i}\, \sum_j \chi_{ij} \,\phi_j    -\sm{N_i}\,\tfrac{1}{2} \sum_{kj} \phi_k \,\chi_{kj}  \phi_j &&&\Big) 
\end{align*} 
Thus in ideally mixing solutions $\chi =0$ with equal molar volumes $\nu_i = \nu_j$, the chemical rate reduces to the standard mass action product of the concentrations  $\xi_{\chi} \sim  V_M \prod_{j=1}^m e^{ a_j \ln \phi_j} \sim V_M \prod_j C_j^{a_j}$ where we used that the volume fractions are proportional to the concentrations by the molar volumes $\phi_j = \nu_j C_j $. 

In general, each reaction thus introduces two free kinetic parameters, the pre-exponential factor and the activation energy of $\kappa$, which
reduce to a single value $\kappa(T)$ under isothermal conditions. 
\subsection{Evaporation}
Consider a thin film, consisting of a mixture containing, among $m-k$ non-evaporating materials (solutes and potentially some crystals), $k$ evaporating solvents $A_i$. We describe the evaporation process as a phase transformation taking place at the surface $S$ separating the condensed film from the gas phase, 
\begin{align*}
    A_i^{(\ell)} \to A_i^{(g)},
\end{align*} 
whereby the $i$-th liquid solvent $A_i$ becomes gaseous and leaves the film.

For the rates, we use one of the semi-empirical expressions derived from the Hertz–Knudsen (HK) model~\ccite{persadExpressionsEvaporationCondensation2016},~\ccite{ronsinPhaseFieldSimulationLiquid2021}. For each solvent $i$, we have
\begin{align}
  \xi_{e,i} =    S(t) \, \tfrac{\alpha}{\sqrt{2\pi \,m_i k_B T}}\,  \left( P_i^*(T)\,e^{\mu_i^{(\ell)}/RT} - P_{i}^{\infty}\right) , \label{eq:Evap}
\end{align}
which yields the expected average particle effusion rate through the surface. We approximately account for the slowdown of evaporation due to a reduction in initial surface area $S_0$ as the film becomes crystalline through the relative liquid-to-total film volume $S(t) \approx S_0 \,\tfrac{V^{(\ell)}}{V^{(\ell)} + V^{(s)} }$. The molecular escape probability $\alpha$ depends only on the flow conditions and surface geometry. It can be considered constant given stationary drying conditions at constant temperature $T$~\ccite{bondMeanEvaporationCondensation2004,schlichting_boundary-layer_2017}. For instance, $\alpha$ can be used to account for the drying conditions encountered during spin coating~\ccite{bornside_effects_1993, meyerhofer_characteristics_2008, van_franeker_controlling_2015} and gas quenching~\ccite{ternesDryingCoatingPerovskite2022}.   
The basic assumption of HK, that the liquid-vapor interface is in equilibrium, i.e., $\mu_i^{(\ell)} = \mu_i^{(g)}$, yields the solvent pressure $P_i^*\,\,e^{\mu_i^{(\ell)}/RT}$ at the film surface. Thereby, $P_i^*(T)$ is the $i$-th solvent vapor pressure and $\mu_i^{(\ell)},\mu_i^{(g)}$ are its chemical potential in the liquid and gas phases, and $R$ is the gas constant. The driving force for evaporation is the pressure difference between the film surface and the environment $ (P_i^*(T)\,e^{\mu_i^{(\ell)}/RT} - P_{i}^{\infty}$). For gas molecules of mass $m_i$, with $k_B$ the Boltzmann constant, multiplying the pressure by $\tfrac{1}{\sqrt{2\pi\,m_i k_B T}}$ yields the rate at which particles cross a unit area of a plane from one side.

 All in all, for evaporation, we consider one drying-condition-dependent adjustable parameter, $\alpha$, and $n(n-1)/2$ pairwise interaction parameters $\chi_{ij}$. 

\subsection{Crystallization}
For the derivation of a crystallization rate, we take inspiration from classical nucleation theory (CNT)~\ccite{kalikmanov_nucleation_2013} and the visualizations from spatial simulation approaches, such as \ccite{ronsinPhasefieldModelEvaporation2020}, but mostly we lean on the commonly employed kinetic approach due to Kolmogorov-Johnson-Mehl-Avrami (KJMA)~\ccite{vansiclenRandomNucleationGrowth1996,blazquez_review_2022}.

\subsubsection{A single crystallization process}
We conceptualize crystallization as a particular phase transition (first order) where an liquid or solid material $(\ell/ s_1)$ transitions to a different solid phase state $( s /s_2)$ through the sub-processes of nucleation and growth. Consider, for instance, the phase transition of a single material $X$:  
\begin{align*}
     X^{(\ell/s_1)} &\to X^{(s/s_2 )} 
\end{align*}
During nucleation, according to CNT, clusters of crystalline material called nuclei form within the mother phase volume $V_M$ due to thermal fluctuations. These nuclei either shrink or, when they exceed a critical size $r_c\, [m]$, become stable nuclei that grow into macroscopic crystals. Thus, crystals begin to form at a characteristic nucleation rate density $\dot{\eta} \, [\#/m^d s]$ - that is, the number $\#$ of supercritical nuclei per unit time and per d-dimensional mother-phase volume. Note that with a composition-dependent nucleation rate density $\dot{\eta}$, the expected first crystal arrival or induction time is already implicitly determined.

During crystal growth, the sizes of the existing crystals increase. With each crystal, we associate a radius
\begin{align}
    r(t,\tau) = r_c (t \geq \tau) + \int_{\tau}^t g(t') \, f(t') \, dt', 
    \label{eq:CrystalRadii}
\end{align}
whereby $\tau [s]$ is the time of birth of the crystal, and $g \, [m/s]$  is the velocity at which the crystal's interface propagates, i.e., the "growth rate". The "impingement function" $f$ accounts for the geometrical confinement of crystals by boundaries due to other crystals that hinder further growth into the directions where crystal interfaces touch. It decreases monotonically from $1$, corresponding to freely growing crystals at the beginning of the process, to $0$ when the volume is fully crystalline and all crystals impinge.  

The crystallization rate can be expressed as the sum of the volume increases due to nucleation and growth:  
\begin{align}
     \xi_{\text{Cr}}     &=  \tfrac{1}{\nu_{X^{(\ell)}}}\Bigg( \, B_d\,r_c^d\,V_M(t) \, \dot{\eta}(t)  \label{eq:ChrysRate} \\
  &\!\!\!\!\!  +  dB_d \dot{r}(t)   \int_{0}^t   \Big(  r_c  + \int_{\tau}^t g(t')  f(t') dt' \Big)^{d-1} V_M(\tau) \dot{\eta}(\tau)d \tau \Bigg) \nonumber
\end{align} 
Here, $B_d = \tfrac{\pi^{d/2}}{\scriptscriptstyle \Gamma(d/2+1) } $ denotes the volume of a unit sphere, with gamma function $\Gamma$ and $d$ the number of linearly independent directions into which crystals grow, termed "growth dimension". 
The expression can be interpreted as follows: The first part of the right-hand side of \cref{eq:ChrysRate} corresponds to nucleation: During a time step $dt$, a number of $ V_M(t) \, \dot{\eta}(t) dt$ nuclei of volume $B_d\,r_c^d$ form in the mother phase volume $V_M$. The second part corresponds to growth: already-present nuclei contribute volume changes according to their time of appearance. The volume change at time $t > 0$ for each of the $\dot{n}(\tau) d \tau = V_M(\tau) \, \dot{\eta}(\tau)d \tau$ crystals born at $\tau$ is equal to $ \frac{d \, B_d r^d(t,\tau)}{dt} = d\, B_d\, \dot{r}(t) r(t,\tau)^{d-1}$. The volume increase due to all crystals born at all times $\tau$ between $0$ and $t$ is therefore equal to $\int_{0}^t \frac{d \, B_d r^d(t,\tau)}{dt} d\tau = d B_d\, \dot{r}(t) \,\int_{0}^t r^{d-1}(t,\tau) \, V_M(\tau) \dot{\eta}(\tau) d\tau$, whereby $r(t,\tau)$ is given by \cref{eq:CrystalRadii}. 

\cref{eq:ChrysRate} constitutes our general crystallization model.  However, to calculate $\xi_{\text{Cr}}$ for a concrete system, we need to specify further the nucleation rate density $\dot{\eta}$ and growth rate $g$ as well as the impingement function $f$ for each crystallization process. Let us now discuss these quantities in detail for a system relevant to perovskites involving multiple crystallization processes. 

\subsubsection{Example including multiple crystallization processes }
Consider three materials $S, X$, and $C$, where $S$ and part of $X$ are liquid $(\ell)$, and $C$ and part of $X$ are crystalline $(s)$. The process system
\begin{equation}
\begin{aligned}
     X^{(\ell)} + 2 \, S^{(\ell)}  & \rightleftharpoons  C^{(\ell)} &&\text{(P0)}  \\
    C^{(\ell)} & \to C^{(s)} &&\text{(P1)}  \\
    C^{(s)} &\to X^{(s)} + 2 \, S^{(\ell)} &&\text{(P2)}   \\
    X^{(\ell)} & \to X^{(s)} &&\text{(P3)}   
\end{aligned}
    \label{eq:reducedSystemTemp}
\end{equation}
includes a direct crystallization (P3) competing with a two-step crystallization (P0-P2). Once $X^{(\ell)}$ combines with $2\,S^{(\ell)}$ into $C^{(\ell)}$ (P0), $C^{(\ell)}$ crystallizes (process P1) and eventually transforms to crystalline $X^{(s)}$, releasing $S^{(\ell)}$ (P2). 
Thus, if we interpret $S^{(\ell)}$ as the solvent and $X^{(\ell)}$ as the dissolved precursor, P3 represents a direct crystallization, and P0-P2 a secondary crystallization of perovskite via solid-state intermediate phase formation of $X^{(\ell)}$ from solution.

\begin{figure}[htbp]
    \centering
    \includegraphics[scale=0.35]{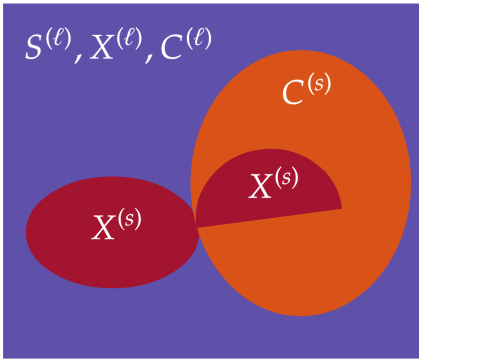}
    \caption{Sketch of material domains for competitive successive crystallization in a dilute liquid phase $(\ell)$. On the one hand, $C^{(s)}$ and $X^{(s)}$ directly crystallize from the liquid phase, and as solids, they occupy separate domains. On the other hand, $X^{(s)}$ resulting from re-crystallization of $C^{(s)}$ appears inside $C^{(s)}$'s domain and $S^{(s)}$ thereby released reenters the common liquid phase.  Because the conditions regarding material supply and crystal impingement differ for the crystallization processes between the liquid phase on the one hand $(\ell)$ and the $C^{(s)}$ phase on the other hand, material supply and crystal impingement need to be distinguished between the processes.}
    \label{fig:InpingmentSketch}
\end{figure}

\subsubsection{Nucleation rate density and growth rate}
To specify nucleation rate densities and growth rates, we draw inspiration from the commonly referred to as LaMer theory \ccite{b.whiteheadLaMers1950Model2021,thanhMechanismsNucleationGrowth2014}. LaMer theory, in essence, introduces two distinct thresholds for nucleation ($\phi_R^c$) and growth ($\phi_R^*$) that determine when each begins and ends, which we refer to as the criticality threshold and solubility limit.    

For each crystallization, we use a reaction volume fraction $\phi_R = \frac{V_R}{V_M}$, where the reactant volume $V_R$ is the amount of material available to the process in the mother phase.
For \cref{eq:reducedSystemTemp} for instance, $\phi_{R_1} = \frac{V_{C^{(\ell)}}}{V^{(\ell)}}$ for P1  and $\phi_{R_2}$ for P2 is one if the mother phase $C^{(s)}$  is available and zero if not. Thus if for example P0 is balanced on the time scale of P1, one effectively has $V_{R_{1}} = \nu_{C^{(s)}} \frac{\kappa_+}{\kappa_-} \, C_{S^{(\ell)}}^{2} \, C_{X^{(\ell)}}$ where $\kappa_+,\kappa_-$ are the forward and backward rates for P0. 

With $\phi_R$ and $\phi_R^*,\phi_R^c$, one can then generally define composition-dependent nucleation and growth rates. In the following, we will, for instance, use  
\begin{equation}
\begin{aligned}
    \dot{\eta} &=    \dot{\eta}_0  \,\max \left( \phi_R - \phi_R^{c} ,0 \right),   \\
    g &=   g_0 \, \,\max \left( \phi_R - \phi_R^{*} ,0 \right),
\end{aligned}
 \label{eq:rateParameterizations}
\end{equation}
where $\dot{\eta}_0,g_0$ are constant scalings. \cref{eq:rateParameterizations} constitutes a simple choice where both growth and nucleation accelerate as the available volume fraction of the crystallizing material increases beyond the thresholds $\phi_R^*$, $ \phi_R^c$. Generally, through refined expressions, additional effects, such as changes in the mixture's diffusion properties, curvature-driven growth, etc., can be accommodated.  

\subsubsection{ Mother phases and impingement functions }
To handle the process system~\cref{eq:reducedSystemTemp}, we introduce the following simplifying assumptions: 
 \begin{itemize}
     \item[A1:] Solid products of a crystallization cannot extend beyond the boundaries of their mother phase.
     \item[A2:] Every crystalline phase can impinge against any other crystalline phase growing in the same phase.
     \item[A3:] Liquid crystallization products end up in a common liquid phase.
 \end{itemize}

In~\cref{fig:InpingmentSketch}, the situation is schematically visualized, assuming that $S^{(\ell)}$, generated by P2, reenters the liquid phase. On the one hand, both C crystals and X crystals, due to the direct crystallizations P1 and P3, are produced from the liquid phase. Therefore, they share the liquid phase as their mother phase and thus can impinge on each other. On the other hand, X crystals due to P2 are produced in the $C^{(s)}$ phase, and therefore their mother phase is the $C^{(s)}$ phase, i.e., a separate solid phase. Thus, if we assume that the growing X crystals due to P2 remain within the $C^{(s)}$ phase domains, they can only impinge on each other but not on X crystals from process P3. 

Therefore, considering~\cref{fig:InpingmentSketch}, we define the impingement volume $V_I$ as the volume of subdomains included in the mother-phase where crystallization is prohibited. 
Based on this definition, we make the following ansatz: 
\begin{align*}
    f  = \phi_I^{\gamma} =   \left(\frac{V_M}{V_M + V_I } \right)^{\gamma}
\end{align*}
Here, $\phi_I$ represents the impingement volume fraction, and \( \gamma \) is a positive empirical exponent. Typically, the mother phase volume $V_M$ is consumed by crystallization, whereas the impingement volume $V_I$ corresponds to the volume of crystalline species. 

Returning to our previous discussion of~\cref{eq:reducedSystemTemp}, for the liquid to solid transitions P1 and P3, the mother phase and impingement volumes are \( V_M = V^{(\ell)} \) and \( V_I = V^{(s)} = V_{C^{(s)}} + V_{X_{P2}^{(s)}} + V_{X_{P3}^{(s)}} \). For the solid to solid transition P2, the mother phase and impingement volume are \( V_M = V_{C^{(s)}} \) and \( V_I = V_{X_{P2}^{(s)}} \), where \( X_{P2}^{(s)} \) and \( X_{P3}^{(s)} \) denote the solid products produced from processes P2 and P3, respectively.

For  processes P1 and P3, the impingement volume fractions are equal and equal to the total liquid volume fraction: $\phi_{I_{1/3}} = \frac{V^{(\ell)}}{V^{(\ell)} + V^{(s)}}$. In contrast, the impingement volume fraction for P2 is given by $\phi_{I_{2}} = \frac{ V_{C^{(s)}}  }{V_{C^{(s)}} + V_{X_{P2}^{(s)}}}$. 
Here, $\phi_{I_2}$ decreases from one to zero - similar to a single process - but with a time-dependent denominator. $\phi_{I_{1/3}}$, however, decreases monotonically from one and reaches zero only if the final state is fully crystalline, which depends on the initial amounts of $2S^{(\ell)}$ and $X^{(\ell)}$.

To determine $\gamma$, we turn to a single crystallization of a pure material $X$ from its liquid phase $X^{(\ell)} \to X^{(s)}$. In this case, one can compare the crystalline volume fraction evolutions predicted by our model, \cref{eq:CrystalRadii} and \cref{eq:ChrysRate}, with those predicted by KJMA. Assuming for this constant rates $\dot{\eta} = \eta_0 \, (t>0), g = g_0\, (t > 0) $ and $r_c =0$, our model is expected to recover $\phi_{\text{KJMA}}^{(s)}(t) = 1-e^{- B_d \dot{\eta} g^d  t^{d+1}}$. Adjusting $\gamma$ to fit our model curves to the KJMA curves, we thus choose the value $\gamma = 0.7$, which we use in the following. \cref{fig:impingement function justification} shows such a fit for some choice of parameters.

Per crystallization, we thus have six parameters that are: The growth dimension $d$, the critical radius $r_c \,[m]$, a growth rate and nucleation rate density scaling $g_0 [m/s], \dot{\eta}_0 \,[\#/sm^d]$, and two thresholds, one saturation limit $\phi_R^*$ per growth rate and one criticality threshold $\phi_R^c$ per nucleation rate density. The latter are, however, obsolete for solid-state transitions ($\phi_R^* = \phi_R^c = 0$).   

\FloatBarrier
\section{Model System for Perovskite film formation}
To illustrate the  typically expected - but not validated-behavior of our general model, for a particular system, we consider coating of the common perovskite reference Methylammonium lead halide (MAPI) from standard solvents such as Dimethylformamide (DMF), Dimethylsulfoxide (DMSO), and N-Methyl-2-pyrrolidone (NMP). For these materials in-situ data on process mechanisms are available~\ccite{stoneTransformationCrystallinePrecursor2018, alhazmi_perovskite_2020,shargaievaTemperatureDependentCrystallizationMechanisms2021,pratapOutofequilibriumProcessesCrystallization2021,reusTimeResolvedOrientationPhase2022,szostakRevealingPerovskiteFilm2020,songRevealingDynamicsHybrid2020,  kneschaurekCompactSampleEnvironment2023, mertenHalideSegregatedCrystallization2024}.

As reported by these studies, perovskite crystallization from a solvent-precursor solution involves several processes: 
solvent ($S^{(\ell)}$) evaporation~\ccite{ternesDryingCoatingPerovskite2022}, solvent and precursor ($\text{AX}^{(\ell)},  \text{BX}_2^{(\ell)} $) coordination into complexes ($C^{(\ell)}$) eventually forming colloidal clusters~\ccite{ pratapOutofequilibriumProcessesCrystallization2021}, and a crystalline
solid state precursor ($ D^{(s)}$), as well as crystallization of perovskite ($X^{(s)}$), either directly or from the solid state precursor~\ccite{shargaievaTemperatureDependentCrystallizationMechanisms2021}. 

Based on this, we propose the following reference process system template for perovskite formation from solution:  
\begin{equation}
\begin{aligned}
{\color{processesAll_Solv}\text{S}^{(\ell)}}  &   \to \text{S}^{(g)}    \\
n\,{\color{processesAll_Solv}\text{S}^{(\ell)}} + {\color{processesAll_Prec}  \text{AX}^{(\ell)}  } + {\color{processesAll_Prec}  \text{BX}_2^{(\ell)}  } & \rightleftharpoons {\color{processesAll_Ic} \text{C}^{(\ell)} }       \\ 
{\color{processesAll_Ic} \text{C}^{(\ell)} }   & \to {\color{processesAll_Ssp} \text{D}^{(s)} }   + \, \sm{m}\, \color{processesAll_Solv}\text{S}^{(\ell)}    \\ 
{\color{processesAll_Ssp} \text{D}^{(s)} }   & \to  {\color{processesAll_Pvk}\text{ABX}_3^{(s)} } + \sm{(n-m)}\,{\color{processesAll_Solv} \text{S}^{(\ell)} }        \\
 {\color{processesAll_Prec}  \text{AX}^{(\ell)}  } + {\color{processesAll_Prec}  \text{BX}_2^{(\ell)}  }   & \to {\color{processesAll_Pvk}\text{ABX}_3^{(s)} }    \\
\end{aligned}
    \label{eq:SystemTemp}
\end{equation}
Although the representation is schematic, we use effective species by summarizing the complete solution chemistry in terms of two chemical reactions. We mostly ignore stoichiometry. This makes the process system generic, enabling it to serve as a starting point for analyzing a wide range of material systems.    
\begin{figure}[htbp]
    \centering
     \includegraphics[scale=0.55]{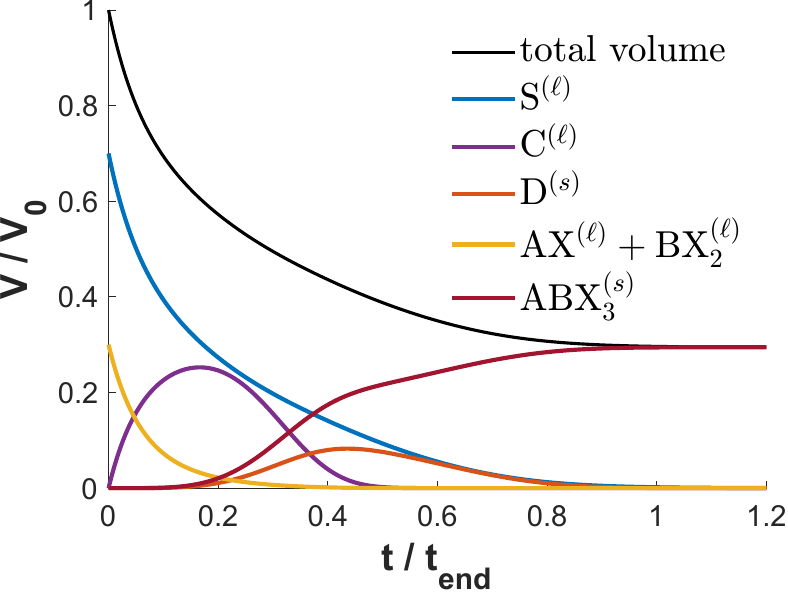}
    \caption{Composition of a drying perovskite ink depending on processing time: $V$ denotes the volumes of the species and $V_0$ the total initial volume. Liquid solvent ({\color{processesAll_Solv}blue}) evaporates reducing the total film volume ({\color{black} black}). Solvent coordinates with liquid precursor ({\color{processesAll_Prec} yellow}) building up complexes ({\color{processesAll_Ic} purple}). The increasingly concentrated complexes accumulate and reorganize into crystalline perovskite ({\color{processesAll_Pvk} dark red}) or solid-state precursor ({\color{processesAll_Ssp} orange}) clusters. Subsequently, the intermediate phase recrystallizes into the perovskite phase, releasing a structurally integrated solvent.}
    \label{fig:All_Processes}
\end{figure}

\cref{fig:All_Processes} shows the corresponding time-resolved composition of a film for an illustrative set of parameters: Dissolved precursor ({\color{processesAll_Prec} yellow}) reacts with the solvent ({\color{processesAll_Solv}blue}), building up complexes ({\color{processesAll_Ic} purple}). Solvent evaporation reduces the total film height ({\color{black} black}) and increases the complex concentration. As the complex concentration rises, perovskite ({\color{processesAll_Pvk} dark red}) or solid-state precursor ({\color{processesAll_Ssp} orange}) starts crystallizing. Once an intermediate phase is present, it recrystallizes into perovskite, thereby releasing the solvent. Finally, when the solvent has evaporated and all intermediate phase has recrystallized, a pure perovskite film remains.

In our modeling framework, the system~\cref{eq:SystemTemp}, with two chemical reactions, one evaporation, and three crystallizations, involves at least $2 + 1 + (6 + 4 + 6) = 19$ free parameters, making parameter studies challenging. Therefore, to understand the basic system behavior, we will focus on the successive crystallization subsystem:
\begin{equation}
    \begin{aligned}
    S^{(\ell)} &\to  S^{(g)}  \\ 
    X^{(\ell)} + 2 \,S^{(\ell)} & \to C^{(s)} \\
    C^{(s)} &\to X^{(s)} + 2 \,S^{(\ell)}
\end{aligned}
\label{eq:redEvapSucc}
\end{equation}
Here, we adapted the stoichiometry to the DMF/MAPI material system \cref{tab:MaterialParameters}. $S^{(\ell)}$ corresponds to DMF, $X^{(\ell)}$ to $\text{PbI}_2 +\text{MAI}$, $C^{(\ell)}$ to $\text{MAPI-2(DMF)}$, and $X^{(s)}$ to MAPI.  
\cref{eq:redEvapSucc} can be viewed as the crystallization sector of our template \cref{eq:reducedSystemTemp}, without direct perovskite crystallization, provided that the chemical equilibrium is balanced on the timescale of evaporation and crystallization.

\FloatBarrier
\section{Model Behaviour}
We turn to studying~\cref{eq:redEvapSucc}. Our strategy is to build understanding from interpretable subsystems, using simulated curves to explore which subsystems — and which curve features — are informative about the process system. The extracted physical interpretations do not depend strongly on the precise parameter values but only on the parameter regimes. For instance, to vary crystallization speeds, we vary the growth rates $g$ rather than the nucleation rate density $\dot{\eta}$. This choice does not affect the interpretations, since increasing either rate leads to more crystal volume appearing at earlier times. 
The simulation results presented are, for concreteness, based on material parameters for Dimethylformamid (DMF) as the solvent and Methylammonium lead halide (MAPI) as the perovskite. For more detailed information on the parameters, see~\cref{app:parTables}.

\subsection{Preliminary study: direct crystallization of a pure material \label{sec:Crys}}
As we use a non-standard crystallization model, we first consider a single crystallization of a material $X$:
\begin{align*}
        X^{(\ell)} \to X^{(s)} 
\end{align*}
Furthermore, we assume that initially all material $X$ is in the liquid state $(\ell)$.

\begin{figure*}[tb]
    \centering
    \labelpanel[scale=0.35]{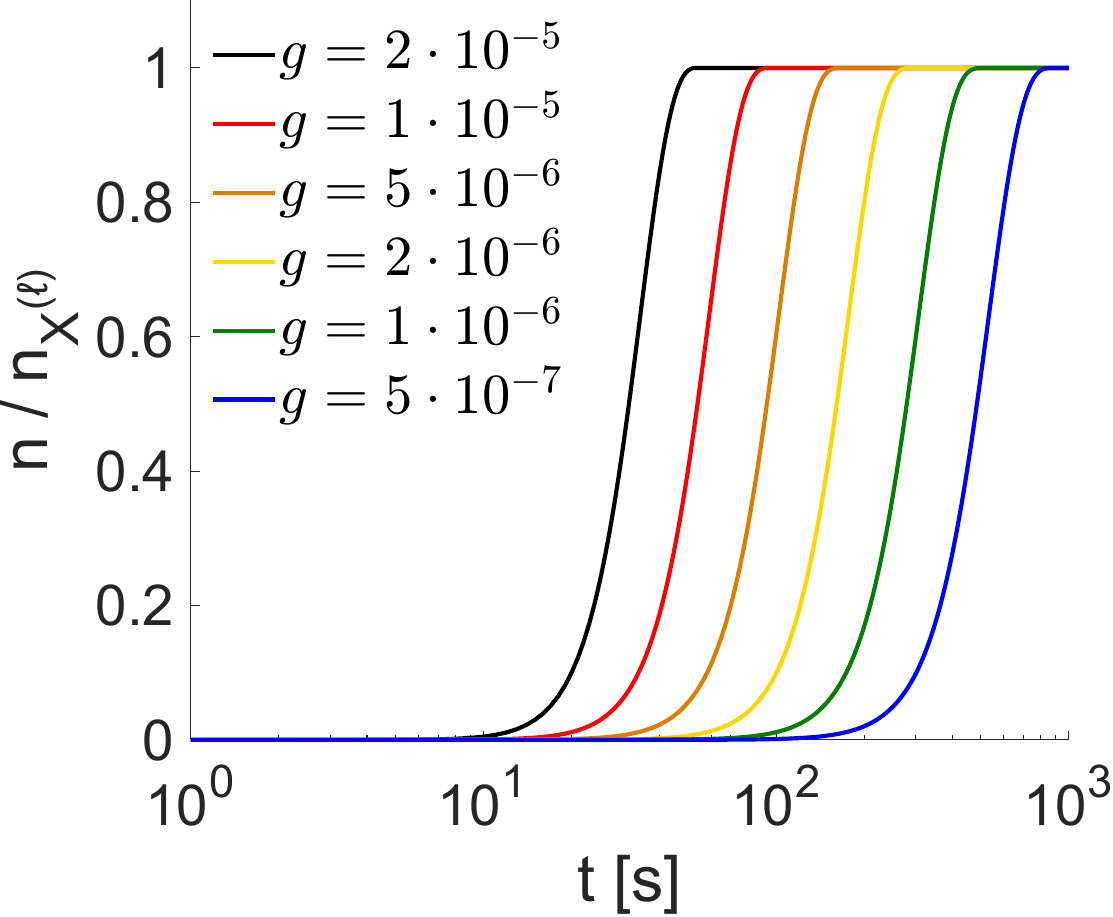}{a}    
    \labelpanel[scale=0.35]{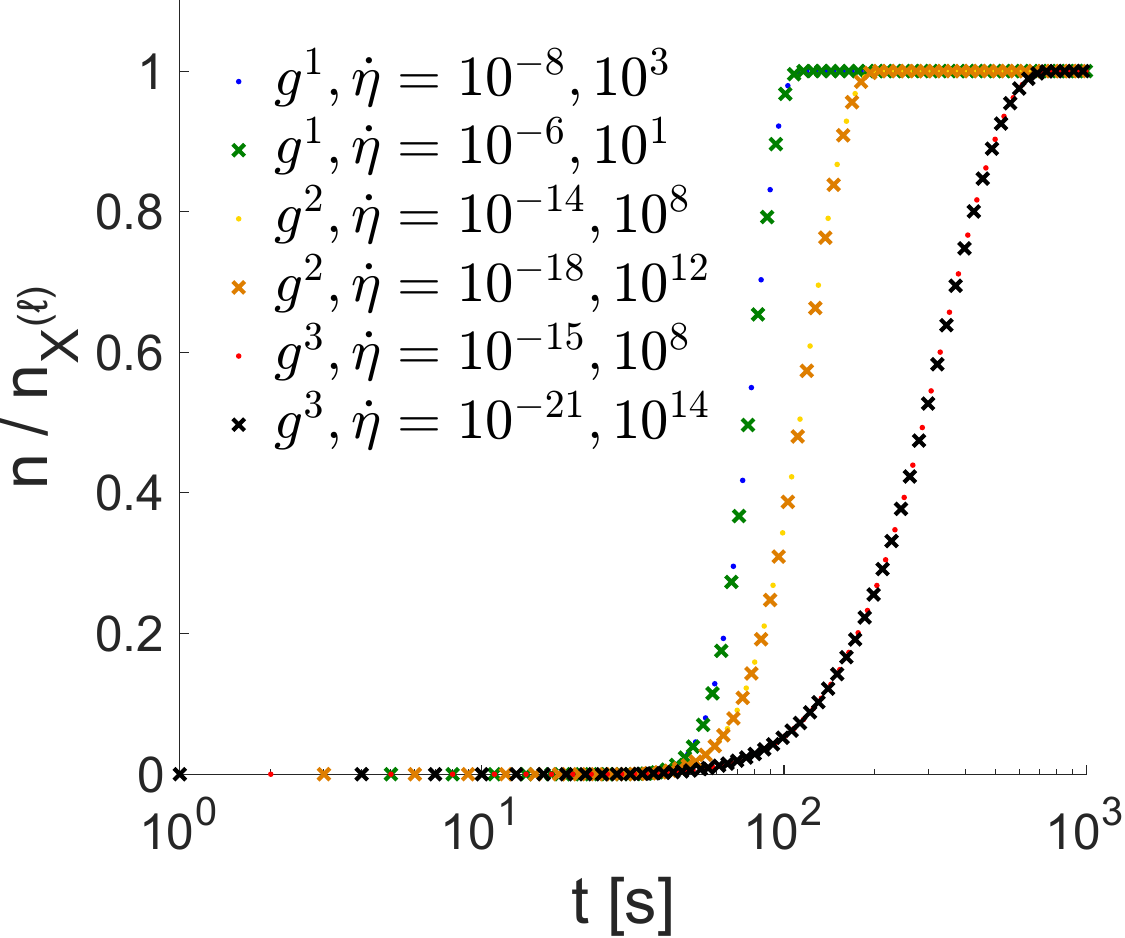}{b}
    \labelpanel[scale=0.35]{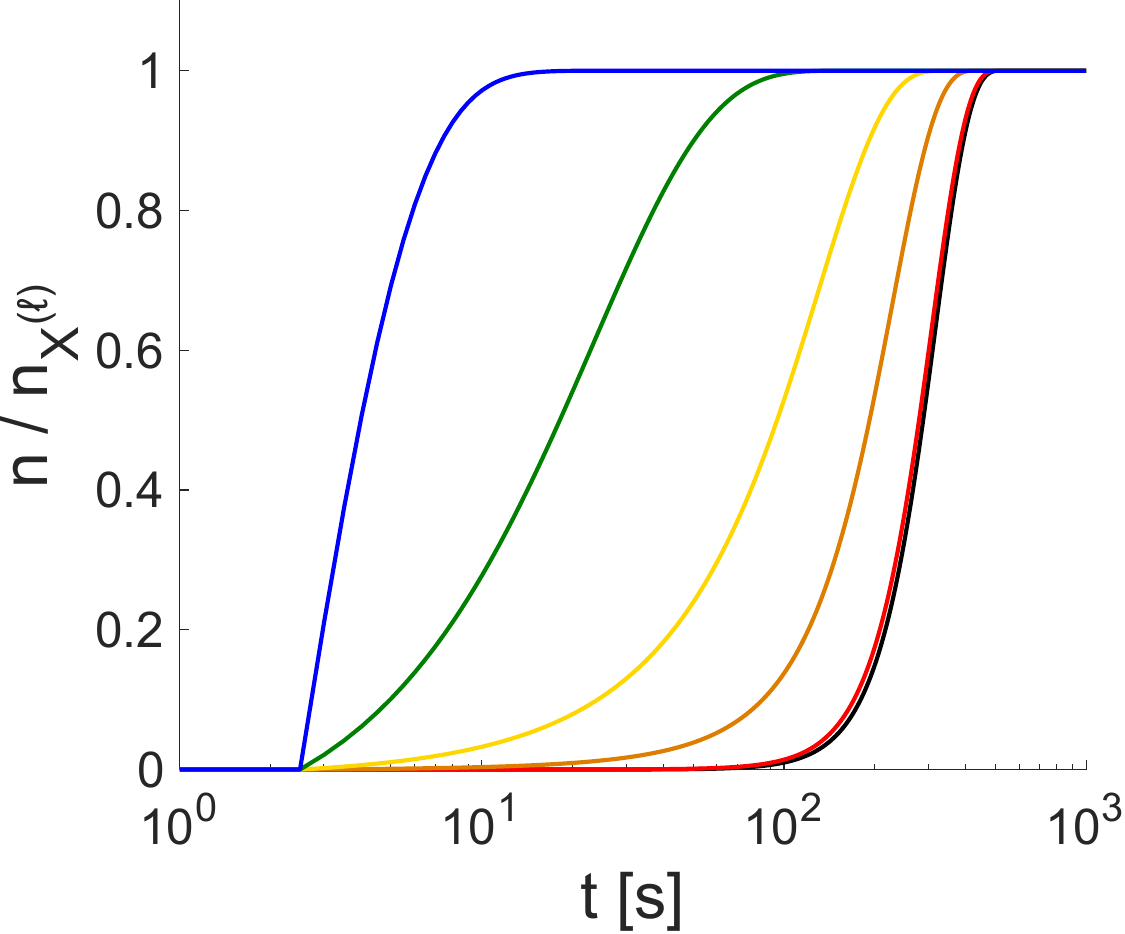}{c}
    \labelpanel[scale=0.35]{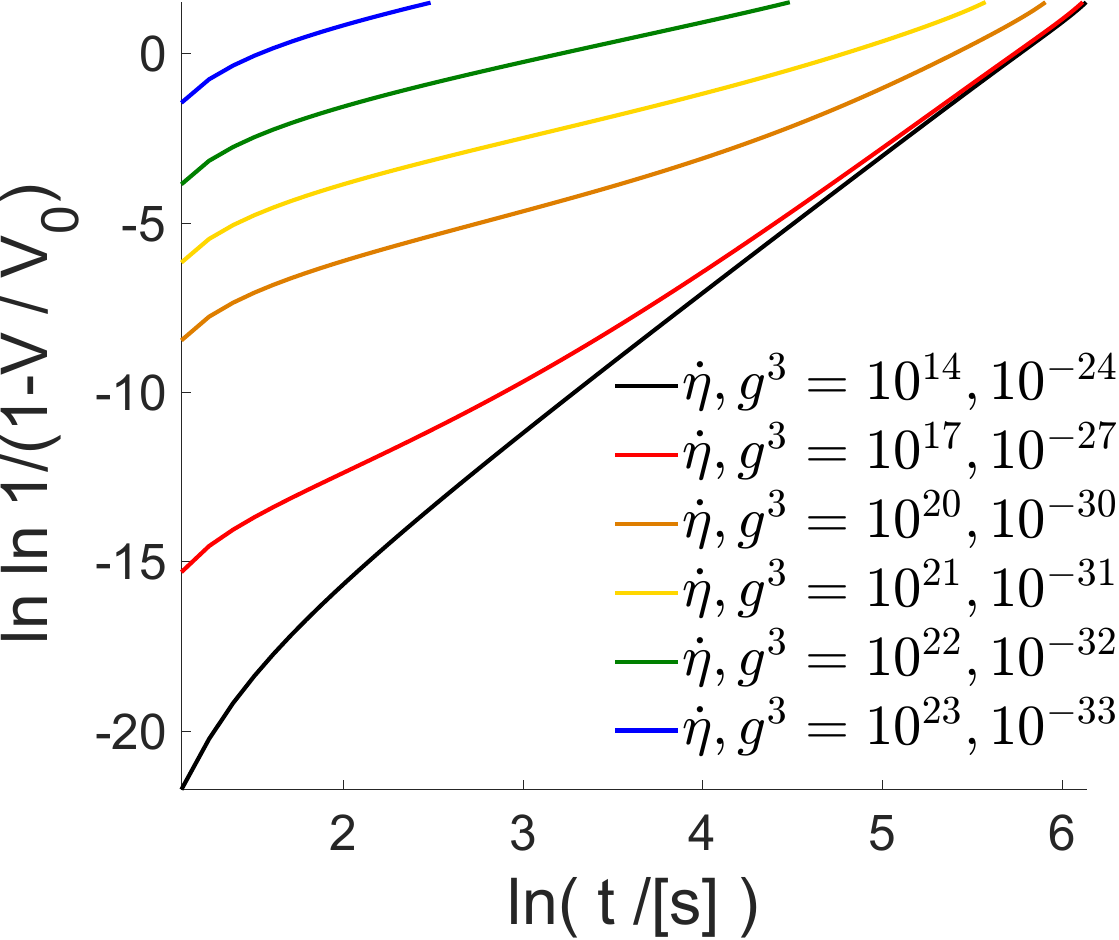}{d} 
    \caption{Crystallization of a pure material: Relative time-resolved mole numbers of $X^{(s)}$ for a) different growth rates $g$ and b) at constant scaling $g^d \dot{\eta}$ with growth rate $g$ raised to the growth dimension  $d = 1,2,3$, and nucleation rate density $\dot{\eta}$. c) and d) Variation of relative importance of nucleation against growth $g^d \dot{\eta} =\text{const}$ for $d =3$ c), relative time-resolved mole numbers of $X^{(s)}$, d) a monotonic transformed version of c) and the volume fraction called Avrami plot. The crystallization curves in a) shift to later times, from black to blue, as the growth rate $g$ increases. Their sigmoid shape is typically expected. b) Several combinations of distinct nucleation rate density $\dot{\eta}$ and growth rate scalings $g^d$ raised by the growth dimension $d$ yield equivalent crystallization curves. This depends on the balance between nucleation and crystallization, as illustrated in panels c) and d). As the importance of $\dot{\eta}$ relative to $g^3$ for $g^3 \dot{\eta} =\text{const}$ increases (black to blue), the Avrami Exponent d) decreases. The degeneracy of the parameter space demonstrated in b) is reflected in the concentration of curves in c) when the growth-rate scaling is dominant.}
    \label{fig:Crys}
\end{figure*}
Consider \cref{fig:Crys}. In panels a) and b), the contributions from nucleated crystalline volume (the first part of the LHS of \cref{eq:ChrysRate}) are negligible ($r_c \ll \bar{r}_{\infty} $). 
The standard sigmoidal shape of the crystallization curves shifts to later times as the growth-rate scaling $g$ decreases. Panel b) demonstrates that the parameter space is degenerate: For each of the three growth dimensions d=1,2,3 (green, orange black) two distinct combinations of growth and nucleation rate density scalings $g,\dot{\eta}$ yield identical curves. That the factor $\alpha_G = g^d \dot{\eta}$ controls the crystallization timescale rather aligns with expectations based on KJMA theory. In contrast, panels c) and d) explore cases where a non-negligible fraction of the crystalline volume has nucleated. In panel c), for $d = 3$ and finite $r_c = 10^{-8} \,[m]$ we adjust the relative importance of $g$ and $\dot{\eta}$ while keeping $g^d \dot{\eta}$ constant. For very large nucleation rates, the volume change due to nucleation becomes dominant, and the crystallization curve is no longer controlled by $\alpha_{G}$. While the curves overlap at higher growth rates for a constant $\alpha_{G}$ (as seen in black and red), as nucleation increasingly dominates (from orange to blue), the curves change shape drastically. This transition between nucleation- and growth-dominated regimes is also reflected in the Avrami plot in panel d), where the slope is often associated with the growth dimension. The slope, however, cannot be interpreted as the growth dimension in general; rather, it ranges from 1 to 3, corresponding to the exponents of the nucleation- and growth-dominated regimes, respectively. For the ensuing investigations, we are primarily interested in the features of the curves related to the process system. Therefore, without loss of generality, we restrict variations in crystallization speeds to the standard growth-dominated regime and vary either $\dot{\eta}$ or $g_0$.

\subsection{Solvent-integrating crystallization \label{sec:SSPf} }
Now consider the primary crystallization step as in \cref{eq:redEvapSucc}:   
\begin{align*}
    X^{(\ell)} + 2 \,S^{(\ell)} & \to C^{(s)}
\end{align*}
Here, when a sufficient amount of precursor $X^{(\ell)}$ is dissolved in the solvent $S^{(\ell)}$, the precursor crystallizes into a distinct solid material $C^{(s)}$ that structurally incorporates solvent. 
In contrast to the previously considered direct crystallization (\cref{sec:Crys}), the transformation now occurs in a precursor-solvent environment. Furthermore, it depends on the relative amounts of the two reactant species $ X^{(\ell)}$ and $2 \, S^{(\ell)}$. To appreciate the qualitative behavior difference, we consider variations of the initial precursor volume fraction $
\phi_{0,X^{(\ell)}}$.
\begin{figure*}[htbp]
    \centering
    \labelpanel[scale=0.35]{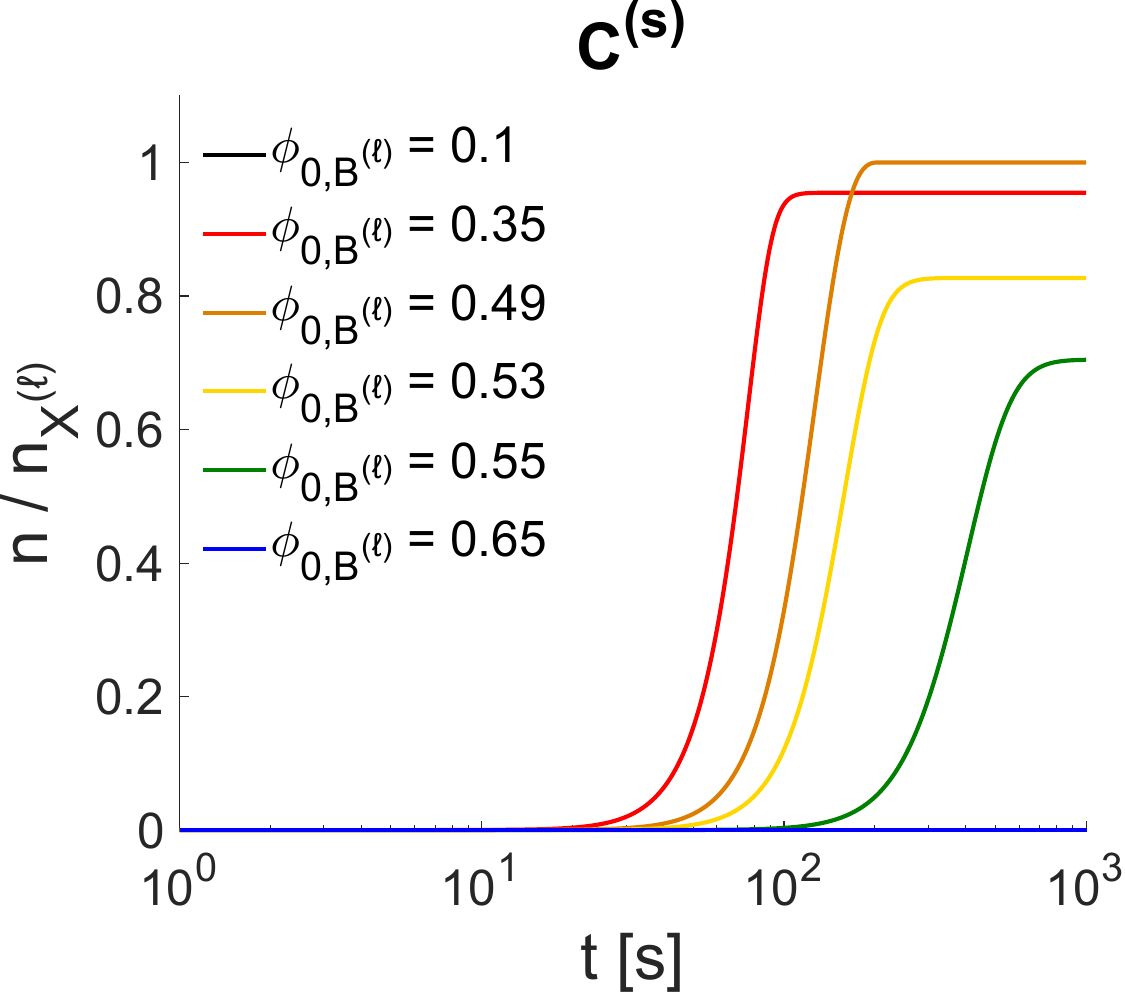}{a} 
    \labelpanel[scale=0.35]{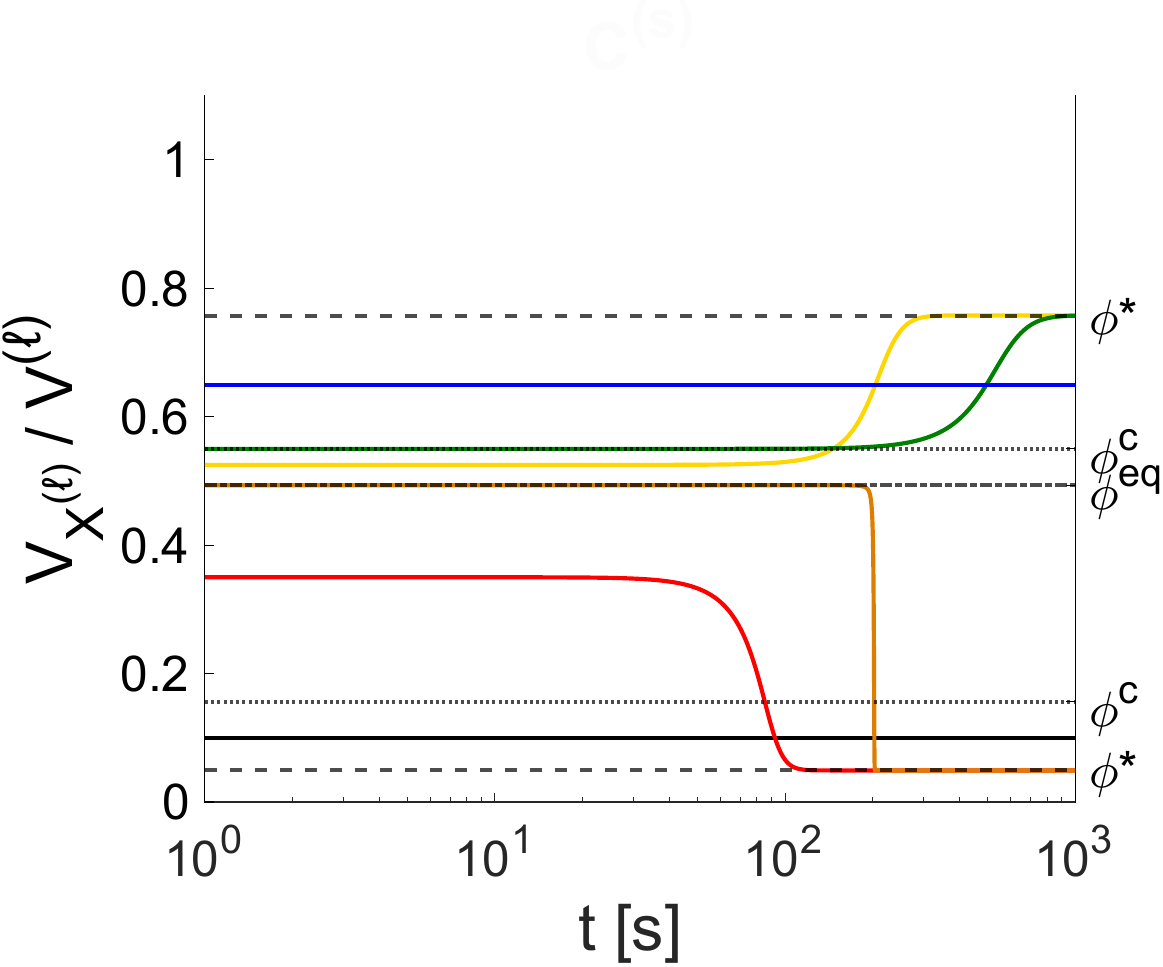}{b} 
    \caption{Solvent integrating crystallization for different initial precursor volume fractions $\phi_{0, X^{(\ell)}}$.  a) Relative mole number of intermediate phase ($C^{(s)}$) molecules against time. b) Precursor volume fraction per time (LaMer Curve) with solubility $\phi^*$ and criticality $\phi^c$ thresholds for precursor ($X^{(\ell)}$) and solvent $(S^{(\ell})$, and equimolar reactant $n_{0, X^{(\ell)}} = 2 \, n_{0, S^{(\ell)}}$ volume fraction $\phi^{\text{eq}}$. The process starts only if the initial precursor concentration $\phi_{0,X^{(\ell)}}$ lies in the critical region between $\phi^c < \phi^{\text{eq}}$ and $\phi^c > \phi^{\text{eq}}$ (between the dotted lines). If the process starts, it proceeds until its supply is exhausted, which occurs when a saturation limit $\phi^*$ is reached. For instance, if precursor is sparse compared to solvent $\phi_{0, X^{(\ell)}} < \phi^{\text{eq}} $ the precursor $X^{(\ell)}$ will be exhausted first and therefore the lower solubility limit $\phi^* < \phi^{\text{eq}}$ is approached.  
    The final amount of intermediate phase $C^{(s)}$ formed therefore is higher for initial precursor volume fractions $\phi_{0, X^{(\ell)}}$ that are closer to the equimolar volume fraction $\phi^{\text{eq}}$ (comparing only volume fractions on the same side, e.g., either below or above $\phi^{\text{eq}}$).} 
    \label{fig:SSPf_VarPhi0}
\end{figure*}

In~\cref{fig:SSPf_VarPhi0} panel a), the transformed precursor mole numbers are plotted against time.
The LaMer graph in panel b) characterizes the instantaneous state of the solution by the time-dependent precursor volume fraction in the liquid phase. The horizontal lines therein correspond to the implied material-specific solubility and criticality limits given by the condition $ \sigma_{N/G}  = k\, \phi_B (1-\phi_B)^2 - \phi_R^{c/*} = 0$. The two resulting limits correspond to the liquid precursor in the solvent $\phi^* < \phi^{\text{eq}}$  and the solvent in the liquid precursor $\phi^* > \phi^{\text{eq}}$. At the equimolar volume fraction $\phi^{\text{eq}} = \frac{ 1\cdot \nu_{X^{(\ell)}}}{1\cdot \nu_{X^{(\ell)}} + 2 \cdot \nu_{S^{(\ell)}}}$, the same number of reactant molecules are available to the process: $n_{S^{(\ell)}} = 2\,  n_{X^{(\ell)}}$. If the initial precursor liquid volume fraction $\phi_{0, X^{(\ell)}}$ lies outside the range set by the criticality limit $\phi^c$ and the equimolar volume fraction $\phi^{\text{eq}}$ — either below it (black, $\phi_{0, X^{(\ell)}} < \phi^c < \phi^{\text{eq}}$), or above it (blue,  $\phi^{\text{eq}}  < \phi^c < \phi_{0, X^{(\ell)}}$) crystallization does not occur. This happens since in both cases one of the two reactants, $2S^{(\ell)}$ and $X^{(\ell)}$, becomes too sparse to trigger nucleation. Otherwise, if the initial precursor volume fraction $\phi_{0, X^{(\ell)}}$ lies within the critical region (between the dotted lines) - $\phi^c < \phi_{0, X^{(\ell)}} <\phi^{\text{eq}}$ (orange and red)
 or $\phi^{\text{eq}} < \phi_{0, X^{(\ell)}} <\phi^c $ (yellow and green) - crystallization occurs. It then proceeds in the direction of the limiting reactant until either the saturation limit above or below the equimolar volume fraction is reached. For instance, if there is less liquid precursor $X^{(\ell)}$ than solvent $2\, S^{(\ell)}$ ($\phi^c < \phi_{0, X^{(\ell)}} <\phi^{\text{eq}}$), the crystallization is limited by the available precursor. The precursor liquid volume fraction approaches the solubility limit of the precursor in the solvent, $\phi^* < \phi^{\text{eq}}$.
For a series of initializations approaching the equimolar reactant initialization (within the critical region) $\phi^{\text{eq}}$ form either below (red, orange) or above (green, yellow), therefore the final amount of crystalline material $C^{(s)}$ increases. The process kinetics, in contrast, depends on the likelihood $\phi_{S^{(\ell)}}^2 \phi_{X^{(\ell)}}$ of finding the three-molecule combination $2\, S^{(\ell)} +  X^{(\ell)} $, the process speed therefore is not necessarily maximal at the equimolar reactant initialization $\phi^{\text{eq}}$.

\subsection{Successive Crystallization of a pure material \label{sec:successive} }
As a first combination of two crystallizations, consider the two successive steps:
\begin{align*}
X^{(\ell)} &\to C^{(s)}  \tag{P$\upbeta$} \\
C^{(s)} &\to X^{(s)}  \tag{P$\upgamma$}
\end{align*}
Here, initially, liquid $X$ crystallizes into solid $C$, which recrystallizes into solid $X$. Thereby, the progress of the first process referred to as P$\upgamma$ depends on the supply of solid $C$ provided by the process P$\upbeta$. Therefore, we will investigate the system's behavior with respect to the relative speeds of the two processes. Without loss of generality, we will vary the speed of P$\upgamma$ relative to the speed of P$\upbeta$, varying P$\upgamma$'s growth rate.
\begin{figure*}[htbp]
\centering
    \labelpanel[scale=0.35]{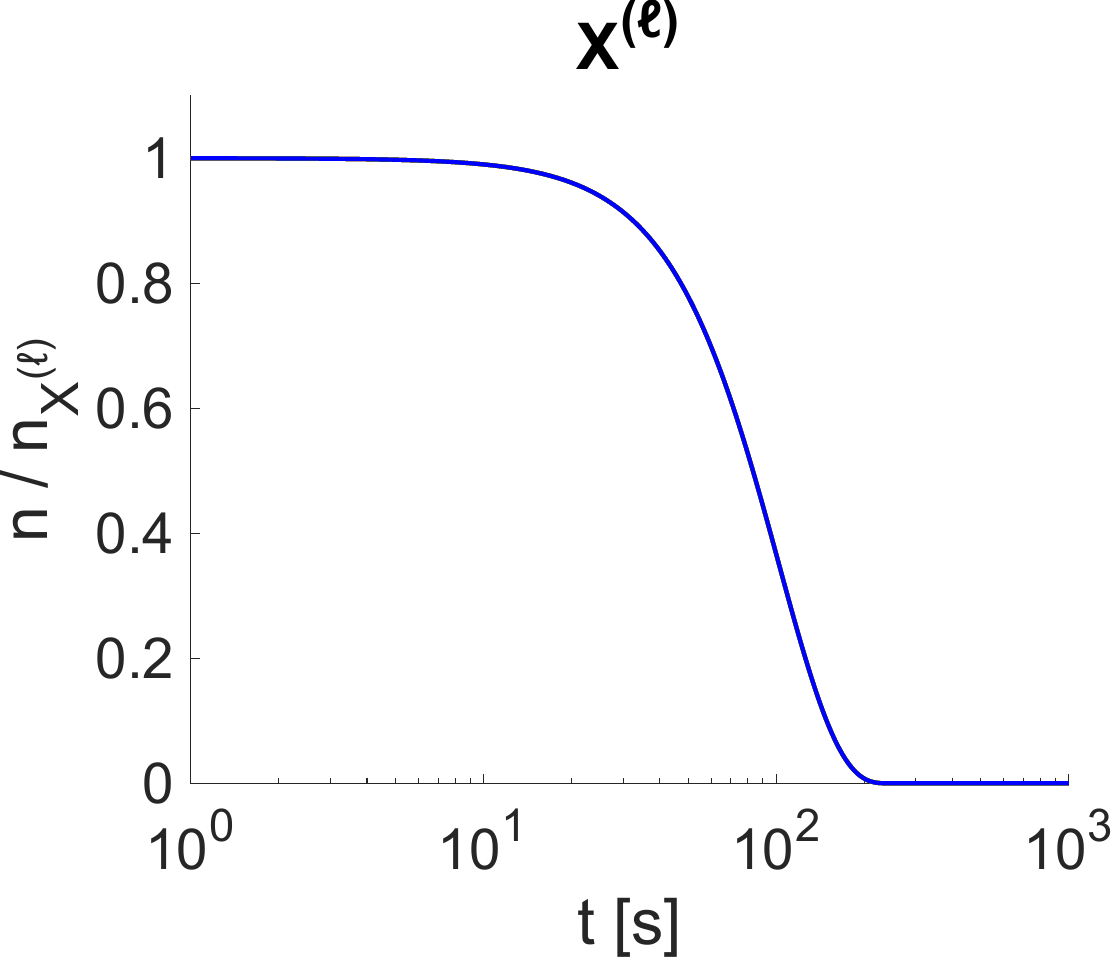}{a}     
    \labelpanel[scale=0.35]{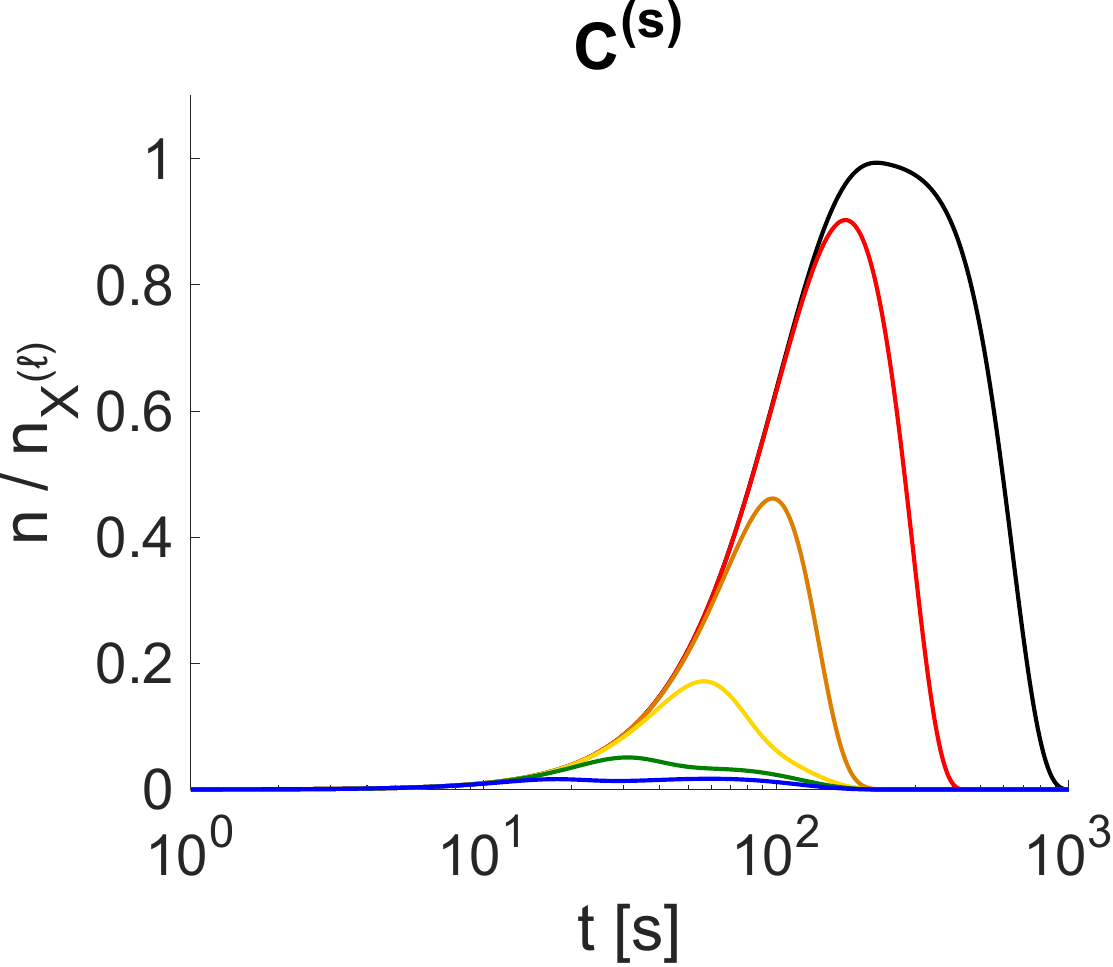}{b}     
    \labelpanel[scale=0.35]{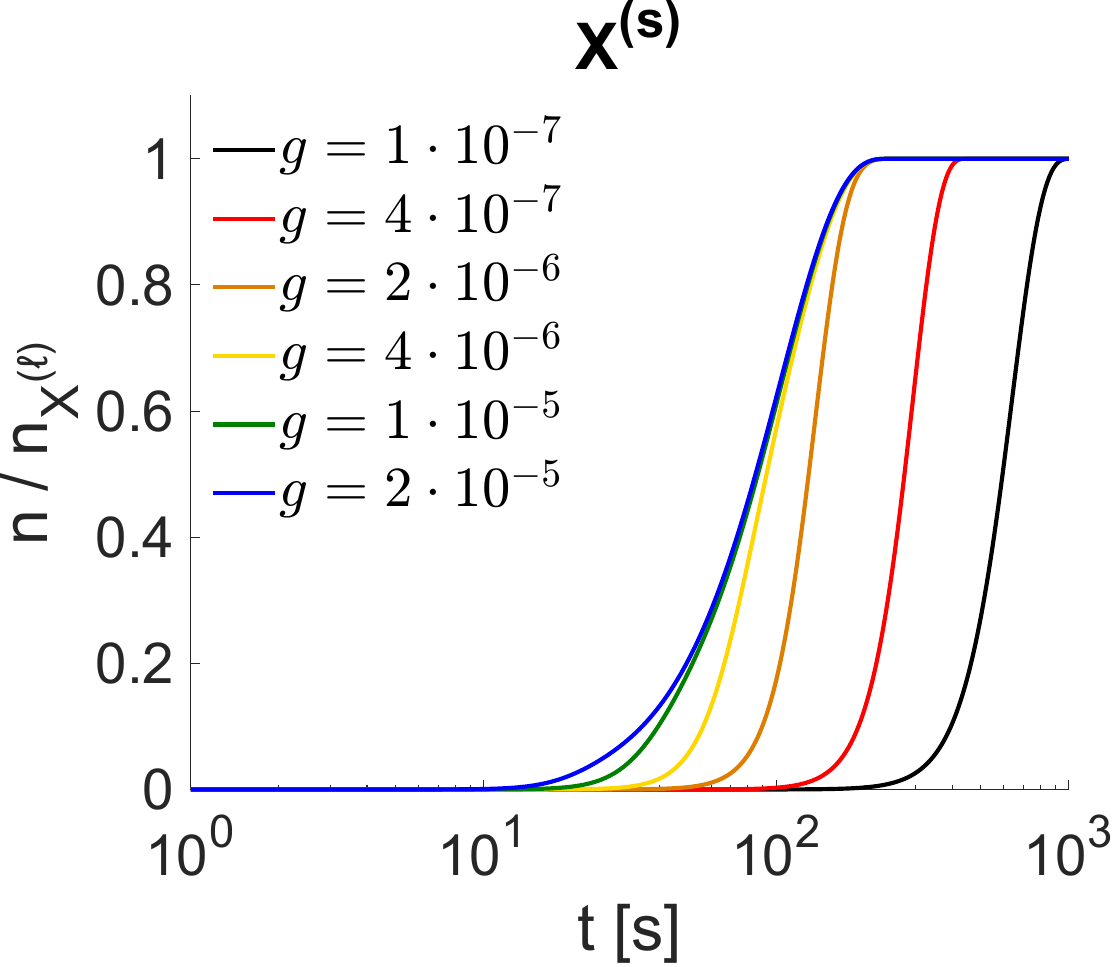}{c}     
\caption{Successive crystallization for different growth rates of the secondary crystallization step. Mole numbers per material relative to the total initial number of moles over time. a) The single collapsed $X^{(\ell)}$ curve (blue) corresponds to the up-flank of the $C^{(s)}$ curves in b). c) As the speed of the $X^{(s)}$ crystallization increases with higher growth rates $g$ (black to blue), the maximal cumulative amount of $C^{(s)}$ decreases. This, in the extreme, leads to a process-limited secondary $X^{(s)}$ crystallization step (yellow to blue).}
\label{fig:Succ}
\end{figure*}
 
In \cref{fig:Succ}, the $X^{(\ell)}$ curve in a) remains unchanged upon variations of P$\upgamma$'s speed. 
If P$\upgamma$ is so slow that the production of $C^{(s)}$ completes before the production of $X^{(s)}$ begins (black to orange), the kinetics of $X^{(s)}$ are independent of the kinetics of $C^{(s)}$. The $C^{(s)}$ curves reflect both the crystallization kinetics of $C^{(s)}$ (the build-up phase) and the crystallization kinetics of $X^{(s)}$. As the speed of P$\upgamma$ increases (from yellow to green), the kinetics of P$\upgamma$ become partly limited by P$\upbeta$. 
Initially, there is an unlimited consumption of $C^{(s)}$ by P$\upgamma$, where it proceeds independently of P$\upbeta$. However, as P$\upgamma$ becomes process-limited by P$\upbeta$, the production of $X^{(s)}$ converges to the up-slope of the $C^{(s)}$ rise. 

In extreme cases, for the green $C^{(s)}$ curve, this can lead to an extended tail after the peak: until P$\upgamma$ becomes limited by P$\upbeta$, the consumption of $C^{(s)}$ by P$\upgamma$ outpaces the production of $C^{(s)}$ by P$\upbeta$. Once P$\upgamma$ becomes limited by P$\upbeta$, its consumption matches P$\upbeta$'s production, resulting in a plateau that eventually ends when P$\upgamma$ stops.

Finally, if P$\upgamma$ is significantly faster than P$\upbeta$ (as shown by the blue curves), the global transformation process is completely limited by P$\upbeta$. In this scenario, the production of $X^{(s)}$ is entirely controlled by the production of $C^{(s)}$. The production of $C^{(s)}$ becomes nearly unobservable, and the observed production of $X^{(s)}$ ultimately reflects the kinetics of $C^{(s)}$ formation. 

More importantly, if the kinetics of $X^{(s)}$ production are entirely dictated by the crystallization of $C^{(s)}$, the absolute height of the $C^{(s)}$ peak will be very low. Therefore, in the case of perovskite formation, the absence of a signal (for example, from in-situ GIWAXS) corresponding to an intermediate phase does not provide sufficient evidence to conclude that formation occurs through direct crystallization. 

\subsection{Successive solvent-coupled crystallization }
As compared to the successive crystallization \cref{sec:successive}, the successive subsystem appearing in \cref{eq:redEvapSucc} and given by 
\begin{align*}
    X^{(\ell)} + 2 \,\,S^{(\ell)} & \to C^{(s)} \tag{P2}\\
    C^{(s)}  & \to  X^{(s)} + 2 \,\,S^{(\ell)} \tag{P3}
\end{align*}
additionally involves the solvent $S^{(\ell)}$.
The release of $S^{(\ell)}$ by P3 therefore always couples the processes, and this feedback from P3 to P2 can alter P2. Therefore, as previously, we will analyze the role of the $S^{(\ell)}$ release by P3 by varying the initial precursor volume fraction.

\begin{figure}
    \centering
    \includegraphics[scale=0.35,trim=0 11cm 0 0]{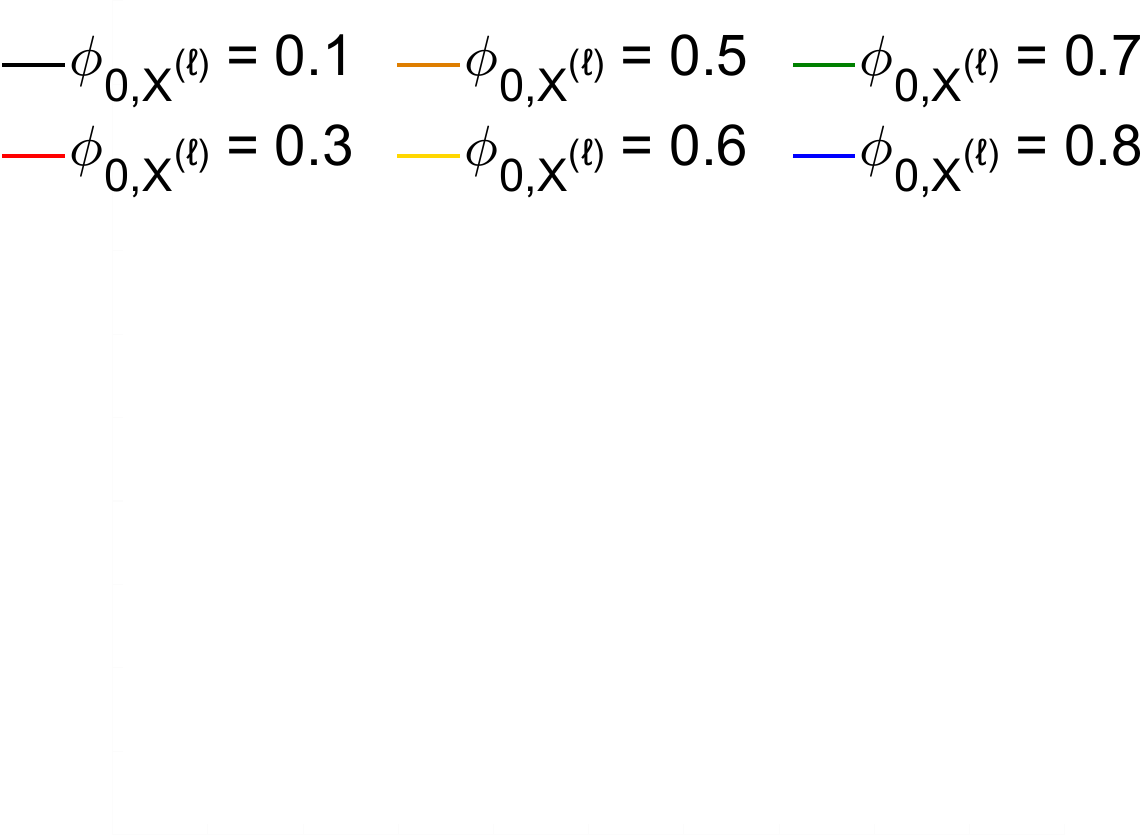}
    \includegraphics[scale=0.35]{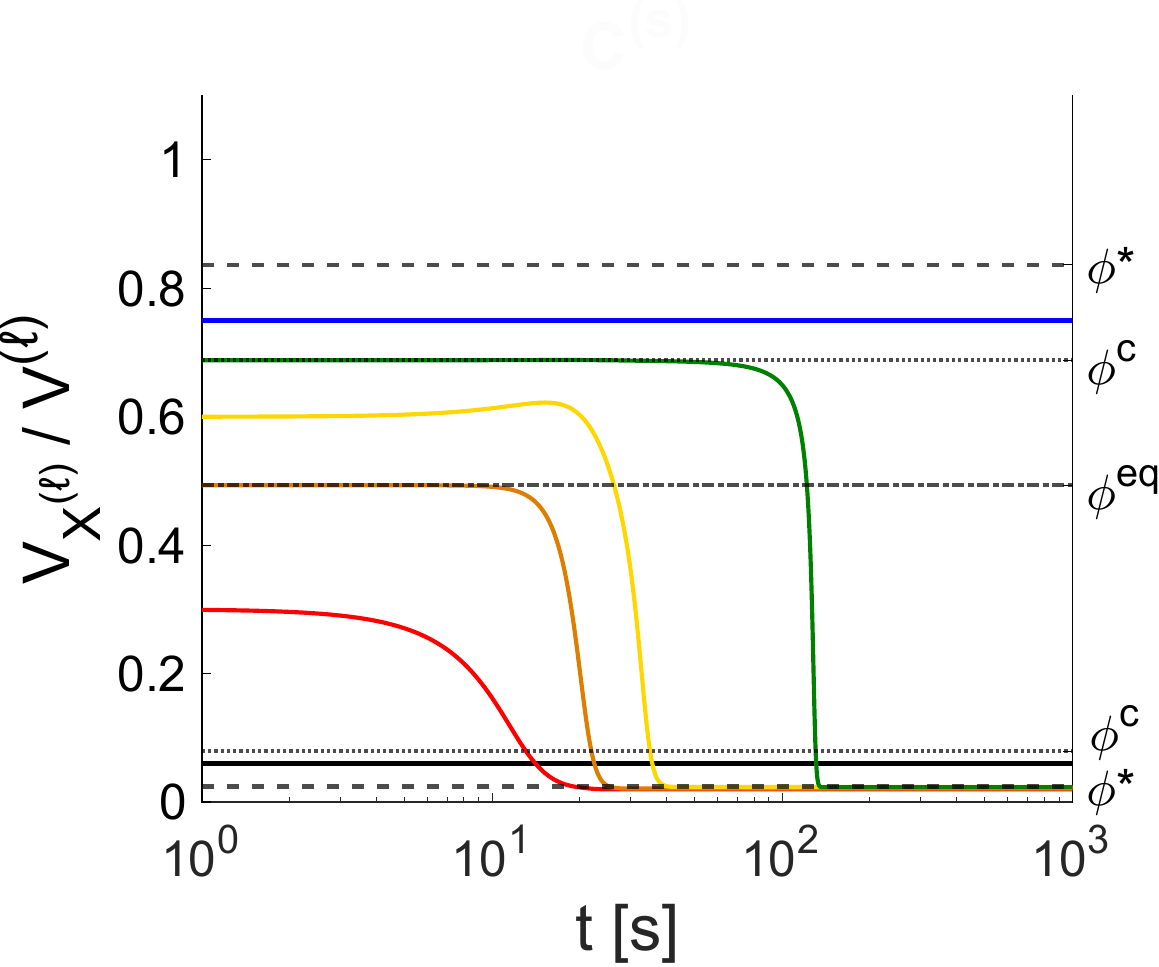}
    \caption{Successive crystallization with solvent coupling for different initial precursor volume fractions $\phi_{0, X^{(\ell)}}$. 
    Liquid precursor volume fraction per time. Crystallization happens only if $\phi_{0, X^{(\ell)}}$ lies in the critical region between the dotted curves defined by the criticality limits $\phi^c$ for the liquid precursor in the solvent and solvent in the liquid precursor, respectively (for green to red but not for black and blue). Crystallization ends only when the precursor saturation limit $\phi^* < \phi^{\text{eq}}$ is reached (lower dotted line) and proceeds towards this limit even starting from initial precursor volume fractions close to the solvent in precursor solubility limit (green).}
    \label{fig:Successsive_Pvkf_VarPhi0}
\end{figure}
\cref{fig:Successsive_Pvkf_VarPhi0} shows the LaMer curve for various initial precursor volume fractions $\phi_{0, X^{(\ell)}}$ (black to blue). The effect of the solvent coupling on the crystallization P2 can be assessed by comparison with  \cref{fig:SSPf_VarPhi0}. If there is an oversupply of either $S^{(\ell)}$ or $X^{(\ell)}$, such that $X^{(\ell)}$ is fully miscible in $S^{(\ell)}$ or $S^{(\ell)}$ is fully miscible in $ X^{(\ell)}$, the process cannot proceed. In contrast to \cref{fig:SSPf_VarPhi0}, however, when $S^{(\ell)}$ is abundant, and $X^{(\ell)}$ is sparse, the process always proceeds to fully exhaust $X^{(\ell)}$. This is possible since P3 resupplies $S^{(\ell)}$. The effect of the solvent coupling, therefore, is that P3 can effectively lift any solvent supply limitation on P2.

P2, as compared to the solvent, integrating crystallization \cref{sec:SSPf}, therefore depends only weakly on the amount of solvent.
Because of this, we expect the material amount curves based on P2 and P3 to show few qualitative differences from those based on the successive system without the solvent coupling \cref{fig:Succ}. To test this, we also vary the speed of the secondary crystallization, P3, relative to that of the first, P2. In particular we consider a fixed initial volume ratio of $S^{(\ell)}$ to $X^{(\ell)}$ within the region where $S^{(\ell)}$ is the limiting reactant $ \phi^c < \phi_{0,X^{(\ell)}} < \phi^{\text{eq}}$.
\begin{figure*}[htbp]
    \centering
        \labelpanel[scale=0.35]{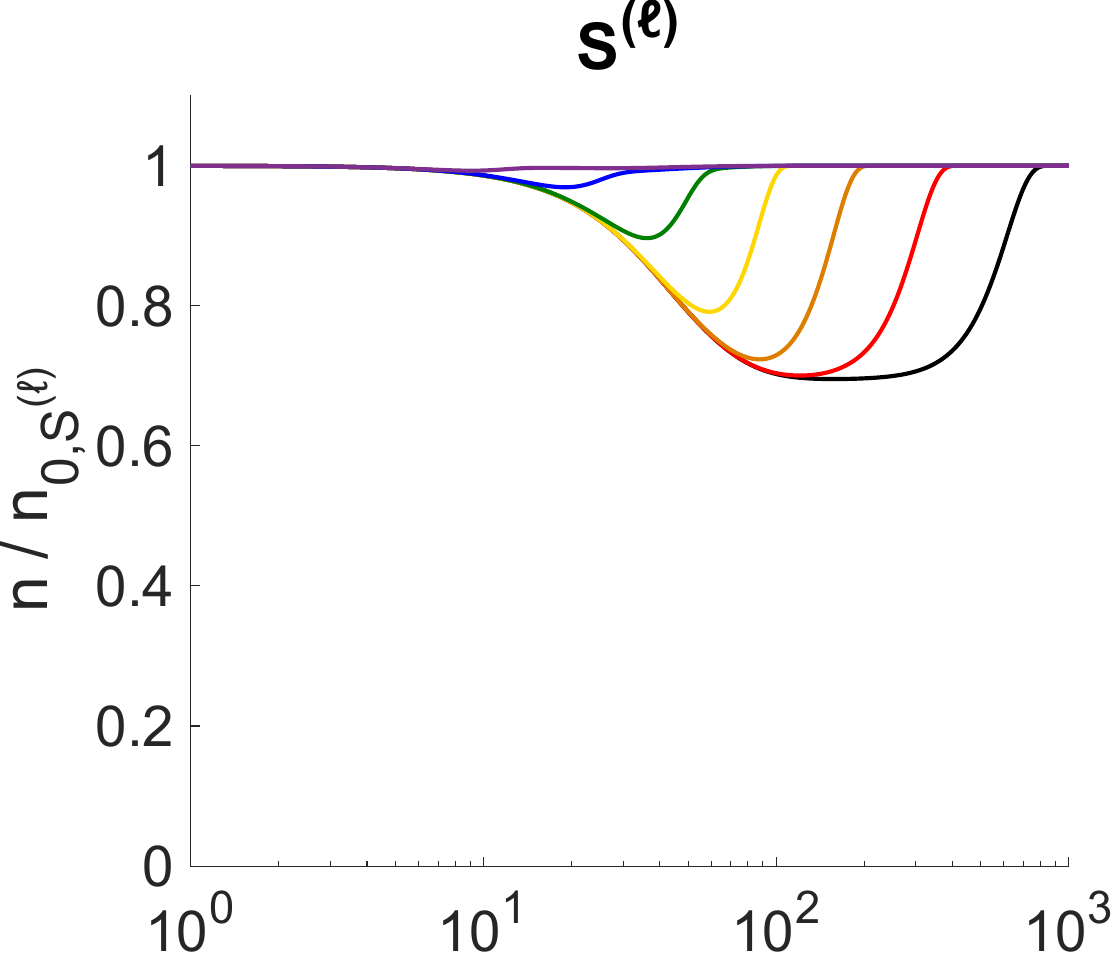}{a}
        \labelpanel[scale=0.35]{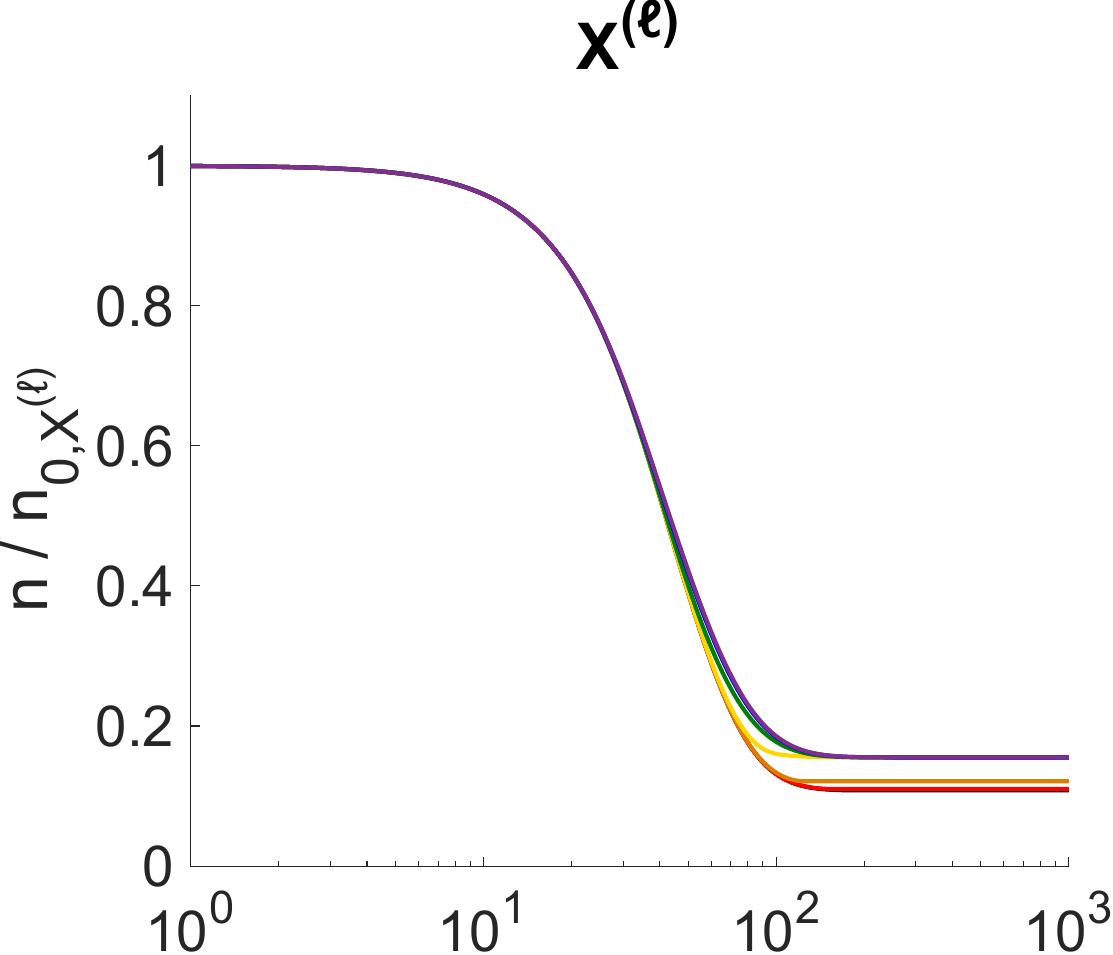}{b}   
        \labelpanel[scale=0.35]{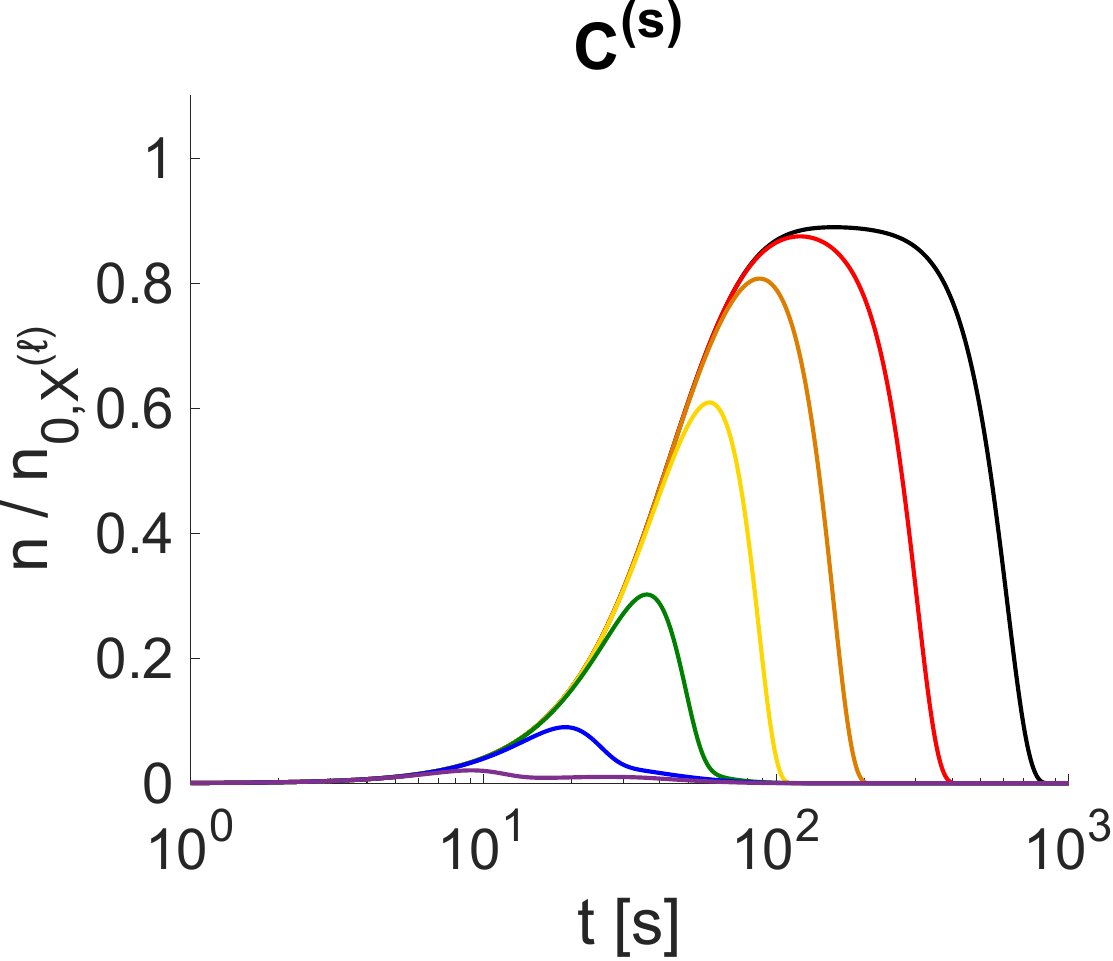}{c}   
        \labelpanel[scale=0.35]{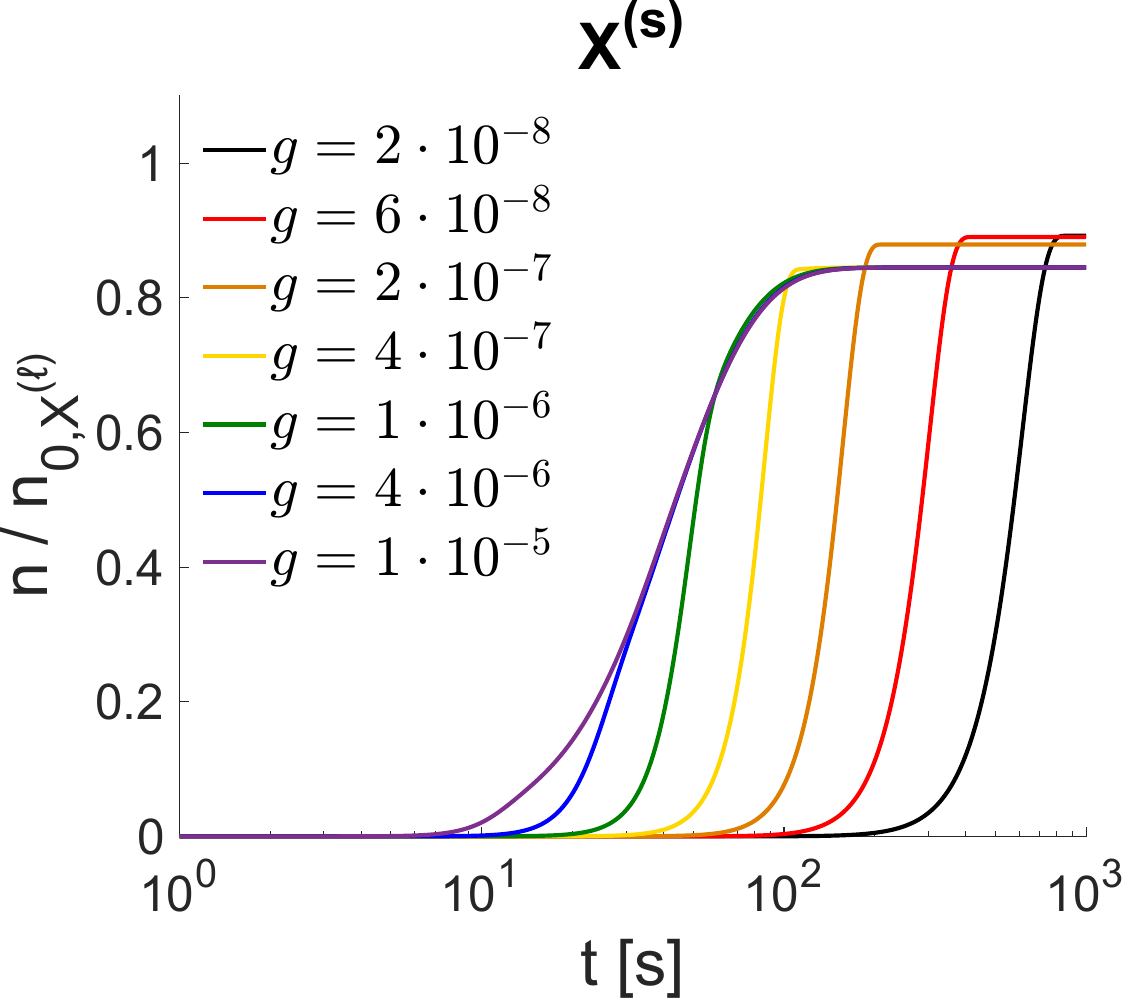}{d}   
        \caption{ Successive crystallization with solvent coupling. Variation of the growth rates $g$ of the $X^{(s)}$ crystallization with fixed solvent-limited $C^{(s)}$ crystallization. Relative material-resolved mole numbers over time. a,b,c) Precursor $X^{(\ell)}$ and solvent $S^{(\ell)}$ are consumed to build up intermediate phase $C^{(s)}$ crystals, which then recrystallize into the perovskite $X^{(s)}$. d) For larger growth rate scalings $g$, $X^{(s)}$ crystallization becomes faster than $C^{(s)}$ crystallization and is limited by it (yellow to purple). For smaller growth rate scalings $g$ (orange to black), $X^{(s)}$ crystallization from $C^{(s)}$ is slower than $C^{(s)}$ crystallization. As initially there are fewer $S^{(\ell)}$ than $X^{(\ell)}$ molecules available for integration into $C^{(s)}$, the crystallization of $C^{(s)}$ is limited in its final stages by the release of solvent from $C^{(s)}$ due to the crystallization of $X^{(s)}$. For the black $C^{(s)}$ curve, this leads to an extended plateau at its maximum height. }
        \label{fig:successsive_Pvkf_Varg2}
\end{figure*}

Compare \cref{fig:successsive_Pvkf_Varg2} with \cref{fig:Succ} based on the successive system \cref{fig:Succ}. Here, in the solvent-coupled successive crystallizations (P2 and P3), even if the secondary step (P3) is slower than the first (P2), the steps do not decouple. This occurs because, initially, there is less solvent reactant $2\, S^{(\ell)}$ than there is of the precursor reactant $X^{(\ell)}$. Consequently, P2 eventually requires P3 to resupply the solvent to proceed (yellow to black).

In the extreme, when P3 is much slower than P2 (black $C^{(s)}$ curve), P2 can become fully P3-limited, yielding a flat $C^{(s)}$ peak. 
Two other differences are evident when comparing the amounts of the final product $X^{(s)}$. First, the $X^{(\ell)}$ conversion is incomplete as the $X^{(\ell)}$ curve does not reach zero.
This is the expected influence of the finite solubility threshold of the precursor in the solvent, which dictates that the precursor can only be consumed above $\phi^* < \phi^{\text{eq}}$.
Second, while the amounts of final product $X^{(s)}$ are constant for the successive system (\cref{fig:Succ}), they can vary for the solvent-coupled successive system (\cref{fig:successsive_Pvkf_Varg2}). One might again suspect the influence of P2's solubility threshold. However, according to \cref{fig:SSPf_VarPhi0}, for a single crystallization from solution that starts at constant initial volume fraction, the amount of final product should not change with the speed of the process. Instead, the differences in the amount of final product arise not from varying initial amounts of solvent, but from differences in the final amount of solvent. If P3 is sufficiently slow (black to orange $X^{(\ell)}$ and $X^{(s)}$ curves), P3 will continue after P2 has stopped because P2's precursor supply is exhausted. As P3 continues to resupply solvent $S^{(\ell)}$, the final solution will therefore contain less precursor $X^{(\ell)}$ than dictated by the solubility limit. Conversely, if P3 is sufficiently fast (yellow to purple), the time lag between the stopping of P2 and P3 becomes small. Consequently, both processes effectively come to a halt when the solubility limit of the precursor in the solvent is reached. The final amount of the perovskite $X^{(s)}$, therefore, again depends only on the initial amounts of the precursor and solvent.  

\subsection{Evaporation-induced, solvent-integrating crystallization \label{sec:evap_SSPf}}
In real-life processing of perovskite films for photovoltaics, crystallization is induced by solvent evaporation. Within our modeling framework, the description of the intermediate phase crystallization thus requires an evaporation process coupled to a single concentration-dependent crystallization:
\begin{align*}
    S^{(\ell)} &\to S^{(g)}  \tag{P1} \\
    X^{(\ell)} + 2 \,S^{(\ell)} & \to C^{(s)} \tag{P2}
\end{align*}
In contrast to the concentration-dependent crystallization discussed in the absence of evaporation (\cref{sec:SSPf}), here the precursor concentration also changes due to the solvent evaporation P1. Evaporation increases the precursor concentration over time. Hence we expect the crystallization P2 to occur eventually - even starting from sub-critical initial concentrations $\phi_{0, X^{(\ell)}} < \phi_c^* < \phi^{\text{eq}}$. Once crystallization begins, its progress depends on the relative rates of evaporation and crystallization. We therefore fix the evaporation-rate parameters while varying the growth rate of the crystallization process, P2.    

\begin{figure*}[tb]
    \centering 
    $     \begin{matrix}
         & \labelpanel[scale=0.35]{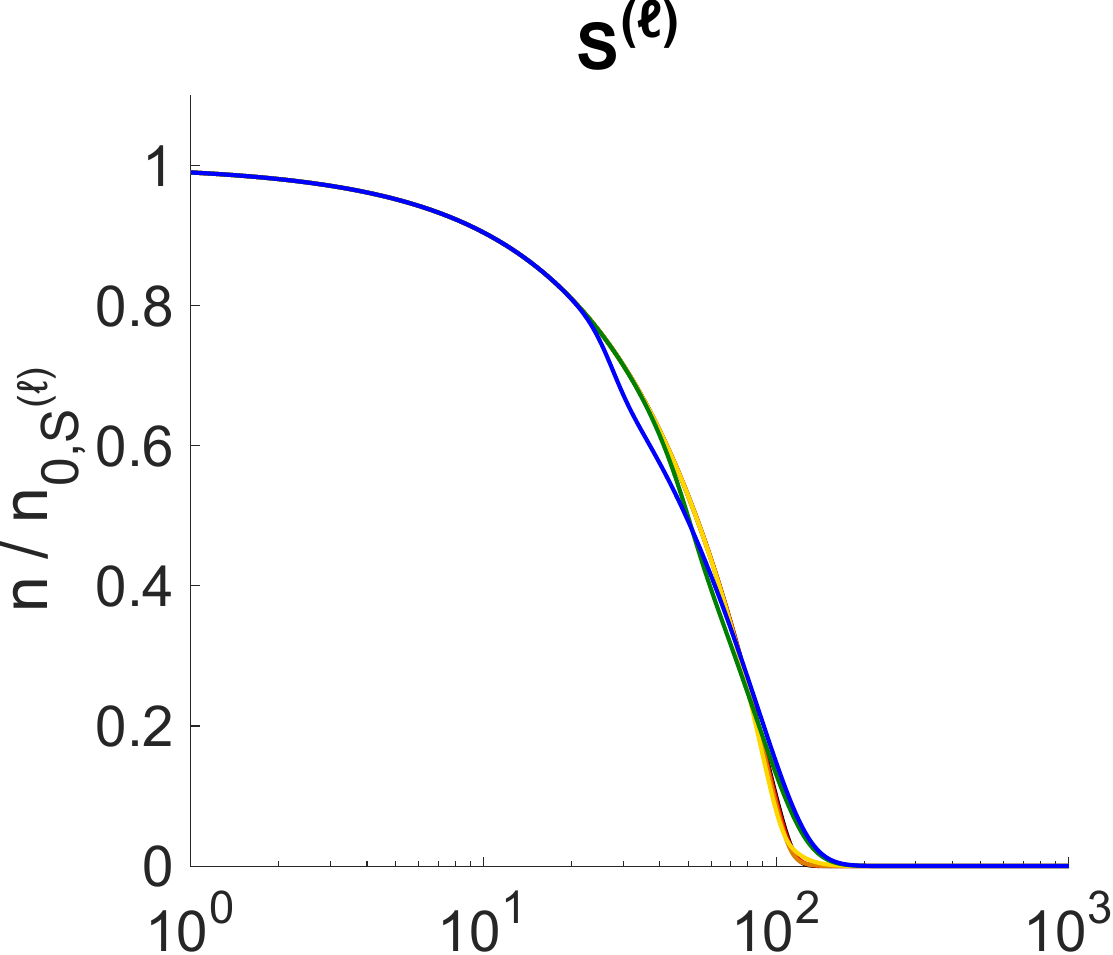}{a}     
         & \labelpanel[scale=0.35]{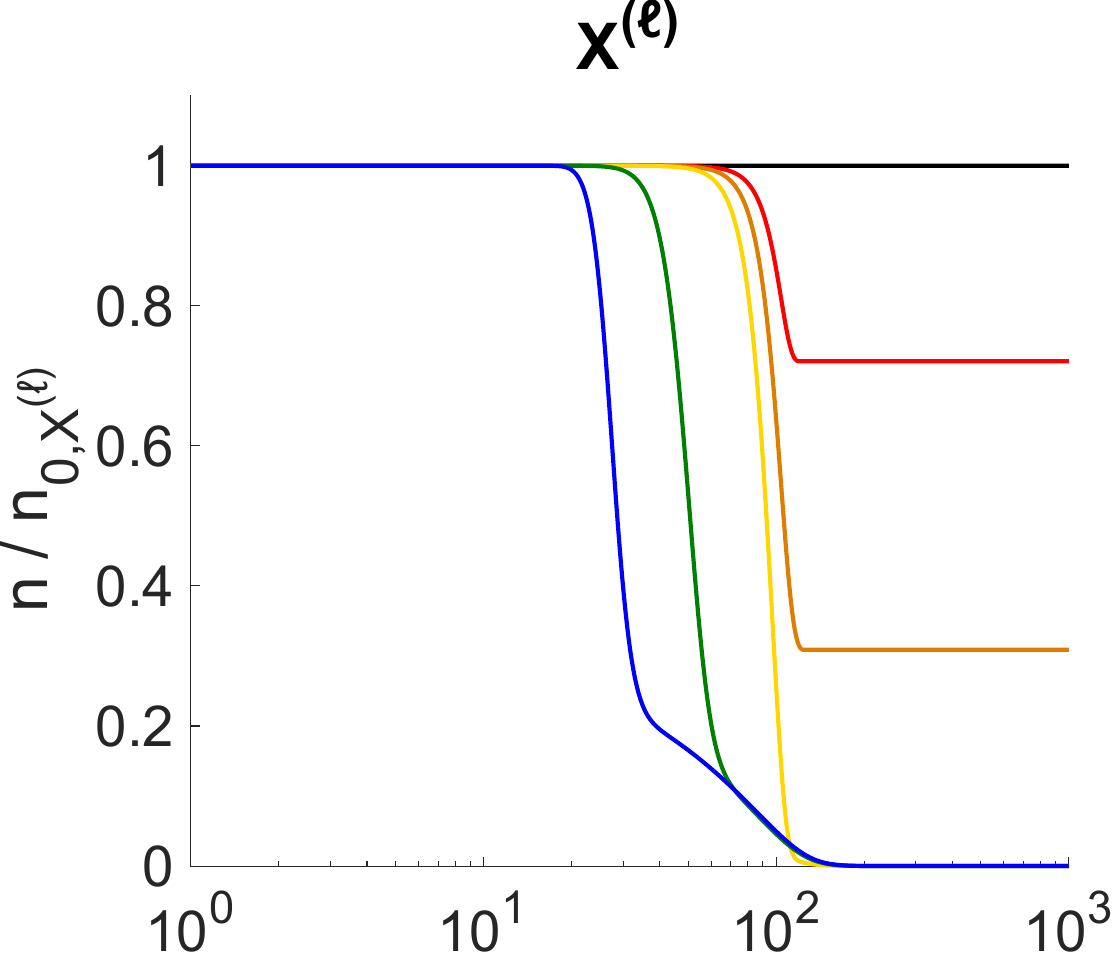}{b}  \\
         & \labelpanel[scale=0.35]{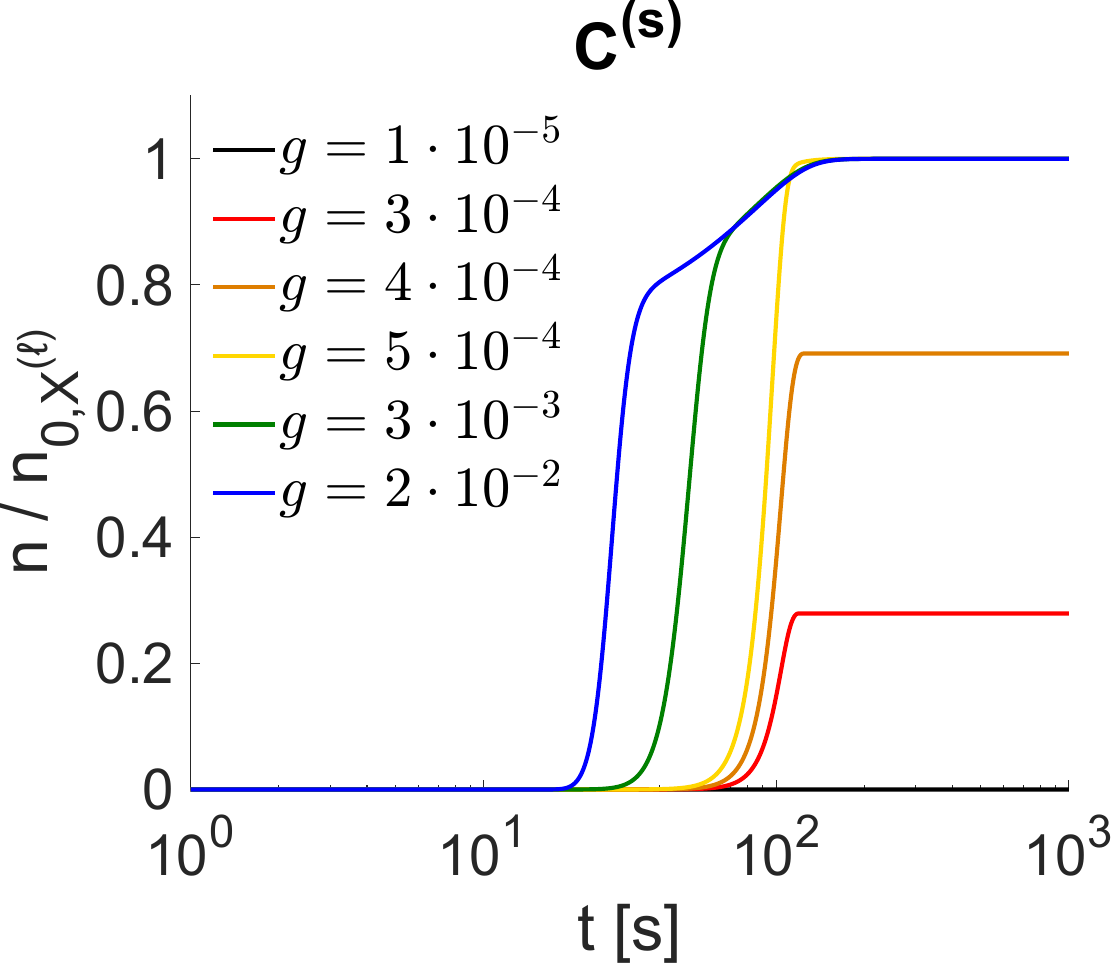}{c}  
         & \labelpanel[scale=0.35]{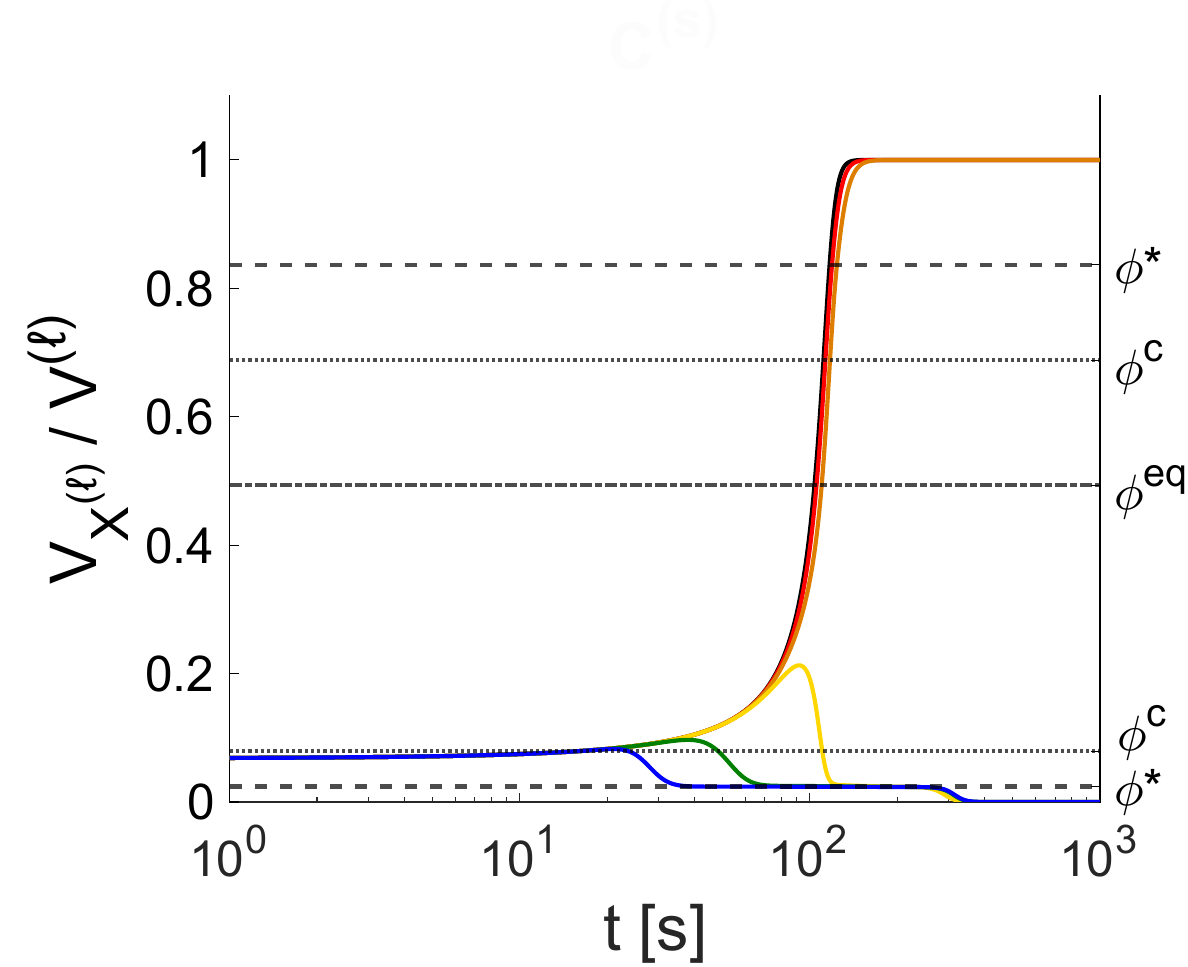}{d} 
    \end{matrix}$ 
        \caption{Evaporation-induced crystallization for various crystal growth rates $g$. a,b,c) Relative material-resolved molecule numbers over time. d) Precursor volume fraction in liquid phase per time. Black horizontal lines indicate characteristic thresholds for solubility $\phi^*$, criticality $\phi^c$ and equimolar reactants $\phi^{\text{eq}}$. a,b,c) $S^{(\ell)}$ evaporation builds supersaturation, inducing $C^{(s)}$ crystallization that exhausts the precursor. c) For large growth-rate scalings $g$ (yellow to blue), crystallization occurs much faster than evaporation. The $C^{(s)}$ crystallization exhausts the supersaturation and then becomes kinetically limited by the evaporation. Otherwise, for small growth rates (orange to black), $S^{(\ell)}$ is completely removed before all $X^{(\ell)}$ is integrated into the $C^{(s)}$, leaving precursor residues (2). b) The LaMer curve correspondingly shows two distinct behaviors (blue to yellow) and (orange to black).       }
        \label{fig:SSPf}
\end{figure*}
Consider the LaMer graphs in \cref{fig:SSPf} panel b). There are two asymptotes corresponding to two final states that we identify from the $X^{(\ell)}$ and $C^{(s)}$ curves: As $S^{(\ell)}$ evaporates, first, the precursor concentration becomes larger than $\phi^{c} < \phi^{\text{eq}}$ and triggers nucleation.
Afterward, if the crystallization is fast (blue to yellow), $C^{(s)}$ buildup proceeds (see $C^{(s)}$ curve) until the precursor volume fraction in the liquid reaches the saturation limit $\phi^* < \phi^{\text{eq}}$. Subsequently, since the crystallization P2 lowers supersaturation faster than evaporation builds it up, P2 is limited by the rate at which evaporation builds supersaturation. This results in a single maximum in the LaMer curve, followed by a constant regime at the solubility limit.
Otherwise, if crystallization is slow  (orange to black), the solvent fully evaporates before the precursor is exhausted, leaving liquid residues of the precursor. 

Both regimes are also visible in the precursor $X^{(\ell)}$ and the intermediate phase $C^{(s)}$ curves. Consider, for instance, $C^{(s)}$: If the crystallization P2 is fast (blue to yellow) after a steep rise corresponding to the free P2 kinetics, a sudden slowdown becomes visible towards the end. This secondary slope in the $C^{(s)}$ curve is entirely determined by evaporation; we therefore say that P2 becomes evaporation-limited. Despite P2's evaporation-limited kinetics, however, in this regime, crystallization proceeds until all precursor $X^{(\ell)}$ available to the process (up to the solubility limit) is finally consumed. In contrast, when crystallization (P2) is significantly slower than evaporation (P1) (orange to black), the solvent can evaporate before the precursor is fully converted to the intermediate phase; therefore, crystallization stops as the solvent is exhausted. 

In summary, slow evaporation relative to crystallization leads to the complete conversion of $X^{(\ell)}$ into $C^{(s)}$ through two subsequent kinetic regimes. In contrast, fast evaporation leads to partial conversion through a single kinetic regime.

\subsection{Evaporation-induced successive solvent-coupled crystallization}
Finally, we add the secondary intermediate phase-to-perovskite crystallization step to the evaporation-induced intermediate phase crystallization system:
\begin{align*}
    S^{(\ell)} &\to  S^{(g)} \tag{P1} \\ 
    X^{(\ell)} + 2 \,S^{(\ell)} & \to C^{(s)} \tag{P2}\\
    C^{(s)} &\to X^{(s)} + 2 \,S^{(\ell)} &  \tag{P3}
\end{align*}

As before, we restrict the $X^{(\ell)}$ volume fraction to below the criticality limit, which ensures that crystallization of $C^{(s)}$ (P2) is evaporation-induced. From our analysis of evaporation-induced $S^{(\ell)}$-integrating crystallization \cref{sec:evap_SSPf}, we expect P2 kinetics to fall into two distinct regimes, depending on its speed relative to the evaporation (P1). 
If the evaporation is slow compared with the crystallization of $C^{(s)}$ ("P1 $<$ P2"), P2 kinetics become evaporation-limited. If evaporation is fast ("P1 $>$ P2"), P2 kinetics become solvent-limited. In addition, the relative speed of the secondary crystallization of $X^{(s)}$ (P3) compared with P1 and P2 may also matter. In principle, each of the six possible speed orderings of P1, P2, and P3 (e.g., "P1 $<$ P2 $<$ P3", "P2 $<$ P1 $<$ P3", ...) could yield distinct behavior. To scan these, we first fix both the evaporation P1 and the $C^{(s)}$ crystallization P2, and then consider the effects of accelerating P3 by increasing its growth rate.

\subsubsection{P2 evaporation-limited  }

\begin{figure*}[htbp]   
    \centering      
    \labelpanel[scale=0.35]{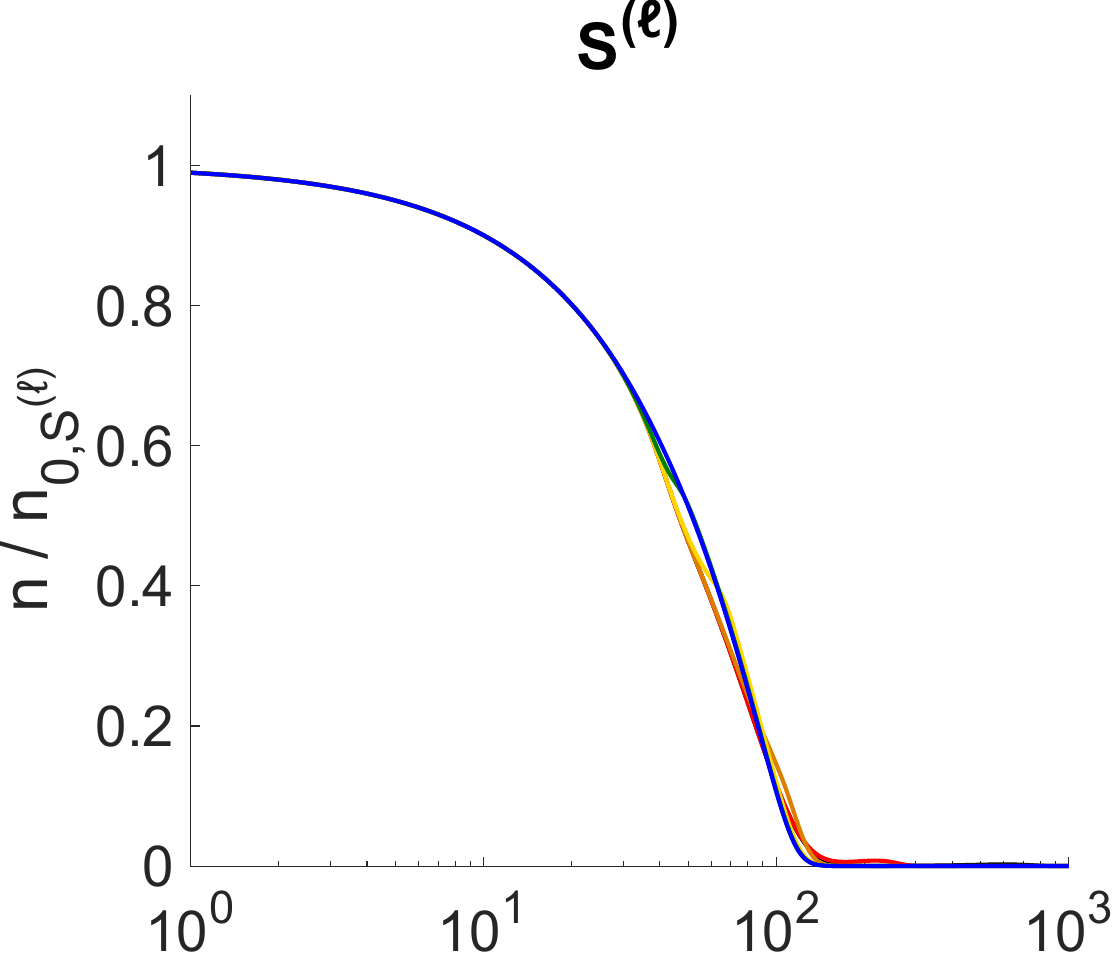}{a}
    \labelpanel[scale=0.35]{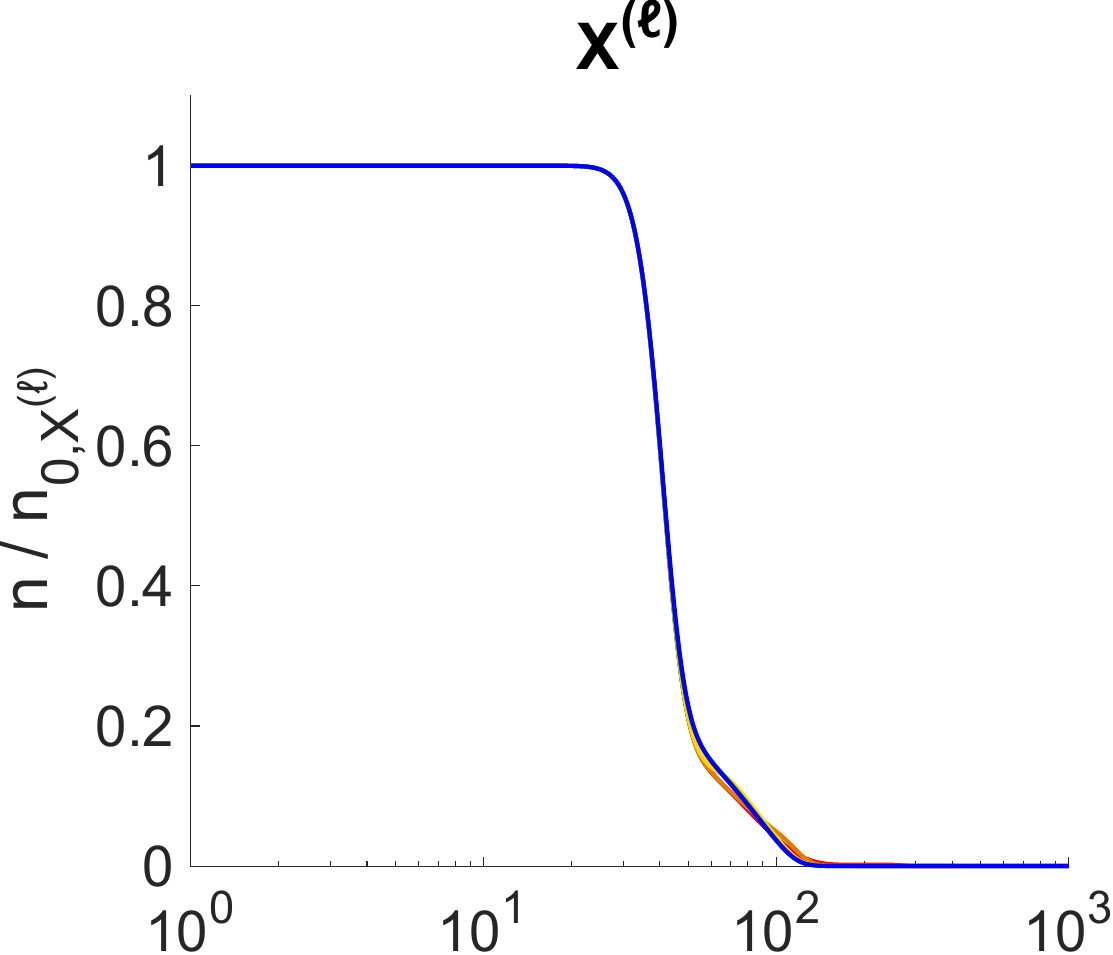}{b}
    \labelpanel[scale=0.35]{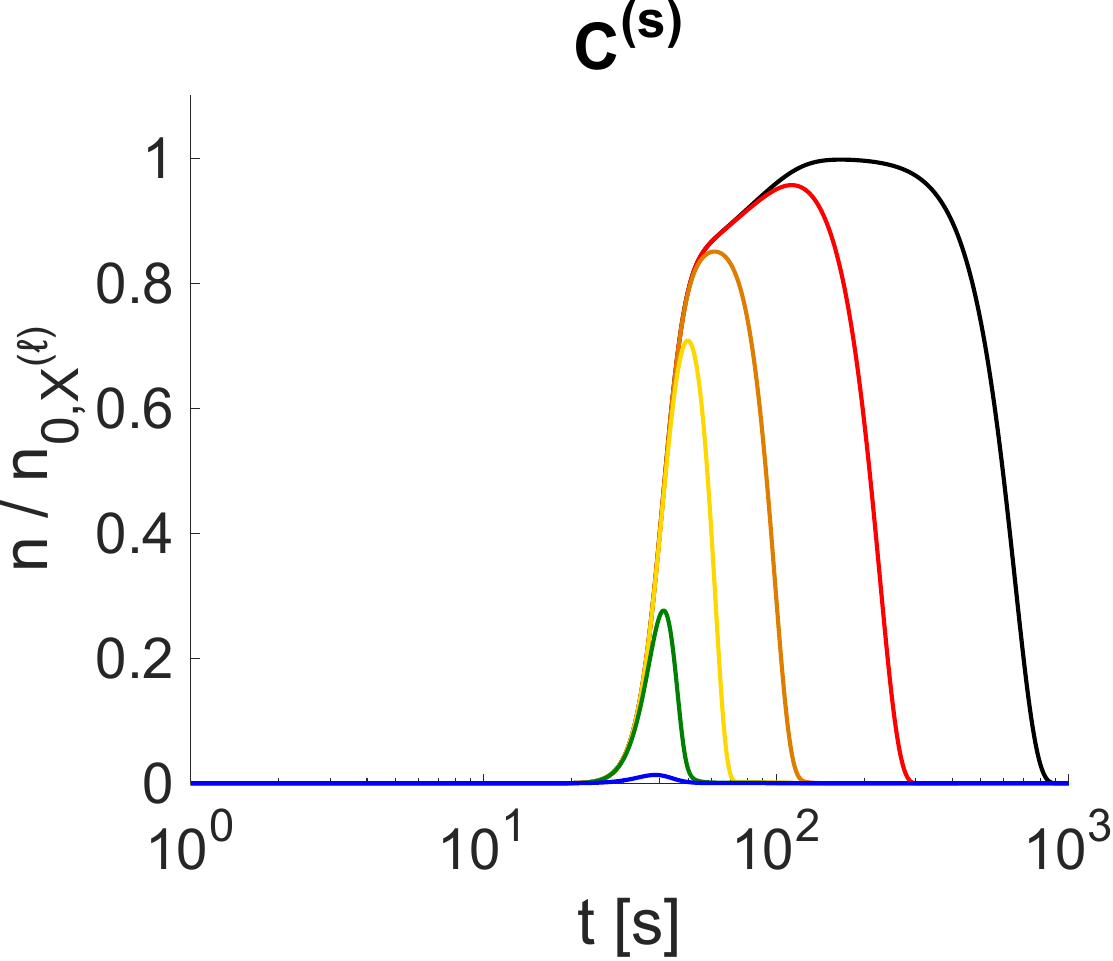}{c}
    \labelpanel[scale=0.35]{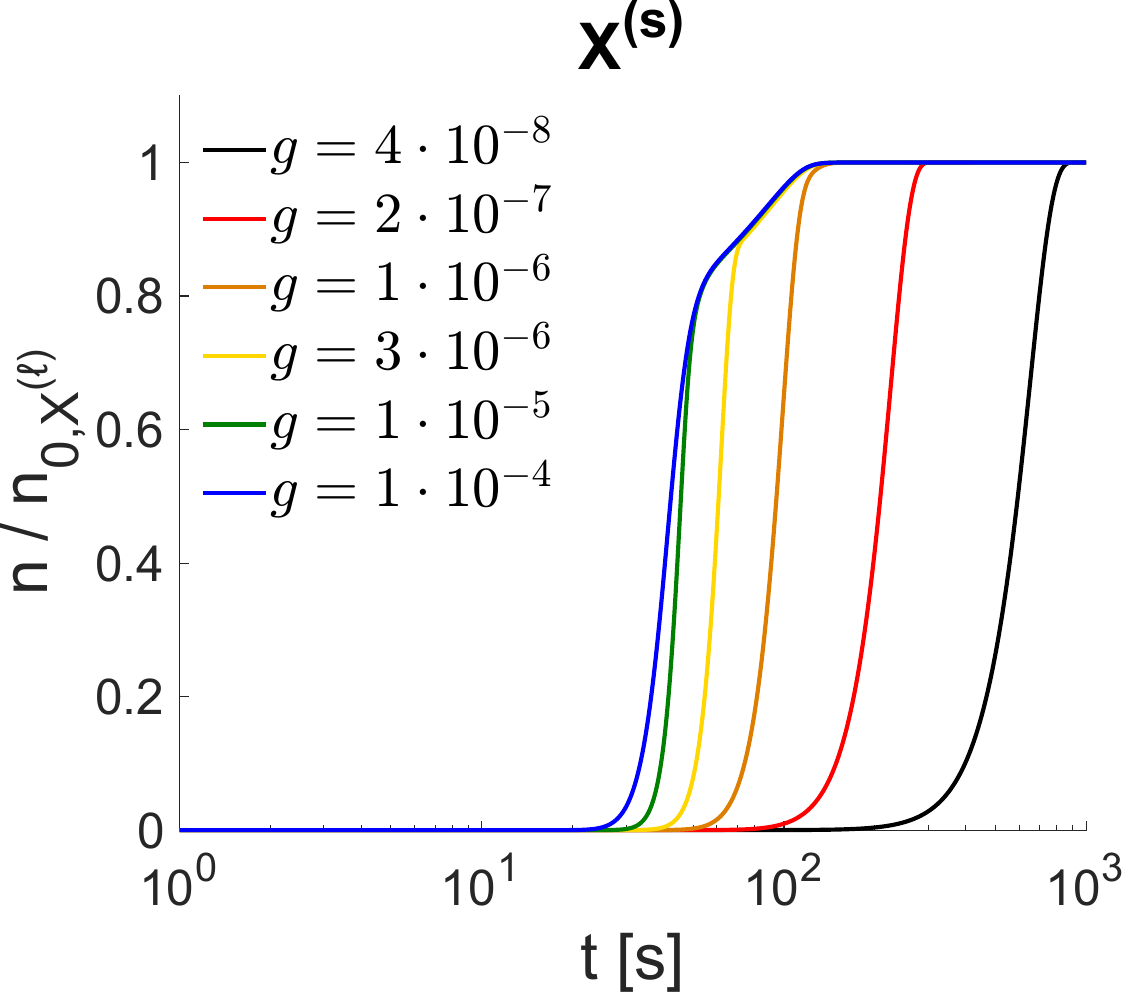}{d}
    \labelpanel[scale=0.35]{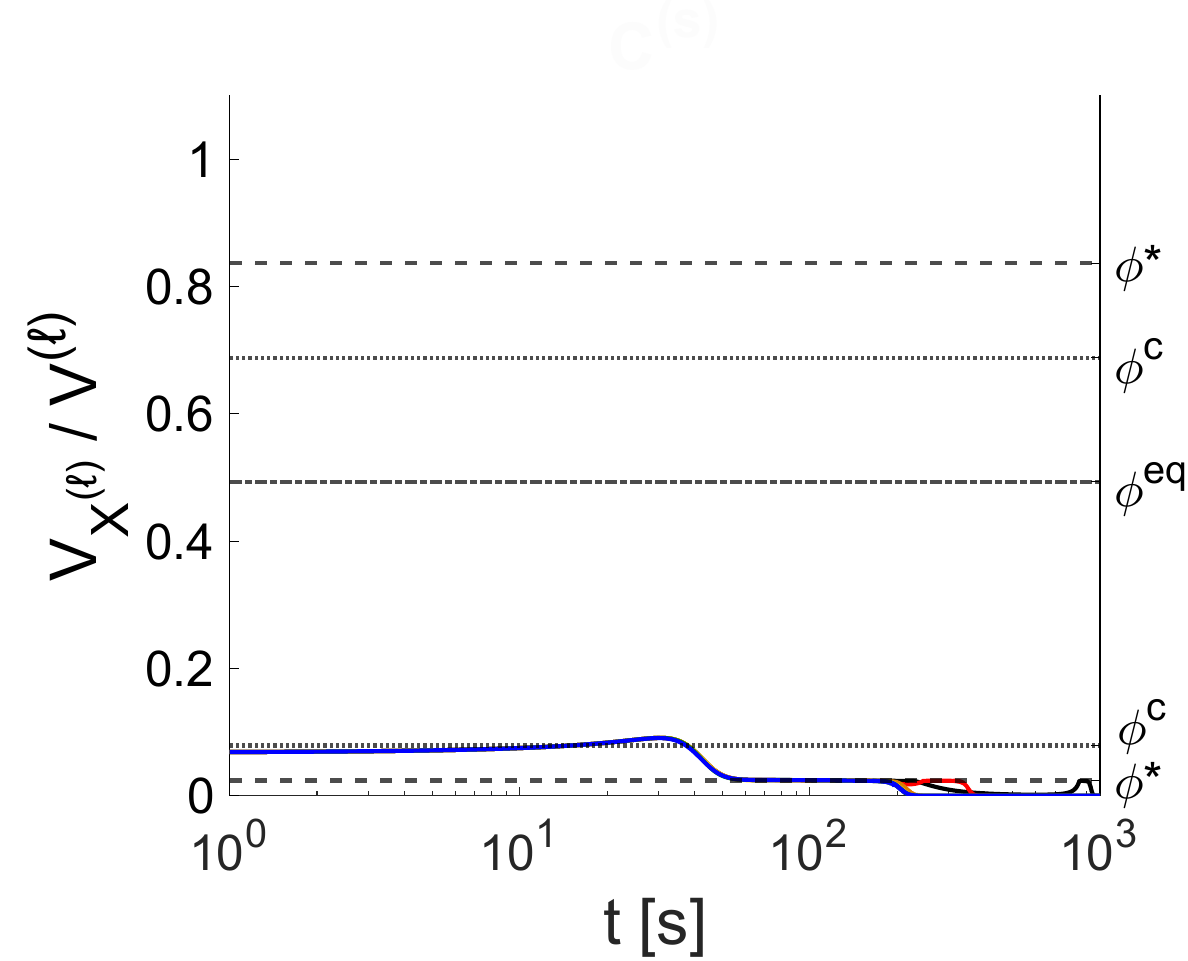}{b}
    \caption{Evaporation-induced successive solvent coupled crystallization for various growth rates $g$ of the secondary $X^{(s)}$ crystallization, starting from an evaporation-induced primary $C^{(s)}$ crystallization that is faster than the evaporation. a,b,c) Relative material resolved time-dependent molecule numbers. b) Precursor volume fractions in the liquid phase. a,b,c) Solvent evaporation induces the $C^{(s)}$ crystallization, which exhausts the precursor $X^{(\ell)}$. c) The $C^{(s)}$ crystallization is limited by the evaporation, as it drains supersaturation faster than it is built up by the evaporation (blue to black). d) Simultaneously, $X^{(s)}$ crystallizes from $C^{(s)}$, and this process can either be limited by the production of $C^{(s)}$ (yellow to blue) or proceed independently (black to orange). In all cases, within the significant crystallization region, the LaMer curves in panel b) collapse to a single curve. Initially, the supersaturation threshold $\phi^c < \phi^{\mathrm{eq}}$ for the liquid precursor in the solvent is reached. Subsequently, it is consumed by the $C^{(s)}$ crystallization, which continues at the saturation limit $\phi^* < \phi^{\mathrm{eq}}$.
    }
    \label{fig:P1eLim_P2VarG}
\end{figure*}

The $C^{(s)}$ curves in panel a) of \cref{fig:P1eLim_P2VarG} show the kinked rise characteristic of evaporation-limited kinetics. This occurs when $C^{(s)}$ crystallization is faster than evaporation, as in the yellow-to-blue curves of \cref{fig:SSPf} panel (a). Likewise, and for the same reason, the conversion of $X^{(\ell)}$ to $C^{(s)}$ is complete in all cases here. The collapsed LaMer curves in panel b) (blue to black) essentially correspond to the single LaMer curve for evaporation-induced-limited crystallization in \cref{fig:SSPf} panel b) (blue and green). 
Therefore, the influence of $S^{(\ell)}$ recycling on the global kinetics is negligible regardless of the relative speed of P3: whether P3 is slower than P2 (black to red), of the same order as P2 (orange), or faster than P2 (yellow to blue).
This might also be expected since $S^{(\ell)}$ is always present in excess. Still, the global $C^{(s)}$ evolution and $X^{(s)}$ formation depend on the kinetic balance between P2 and P3. This manifests as a transition between "kinetics only weakly coupled by $S^{(\ell)}$ recycling" and "P3 fully P2-limited".
\subsubsection{P2 solvent-limited}
\begin{figure*}[htbp]   
    \centering
    \labelpanel[scale=0.35]{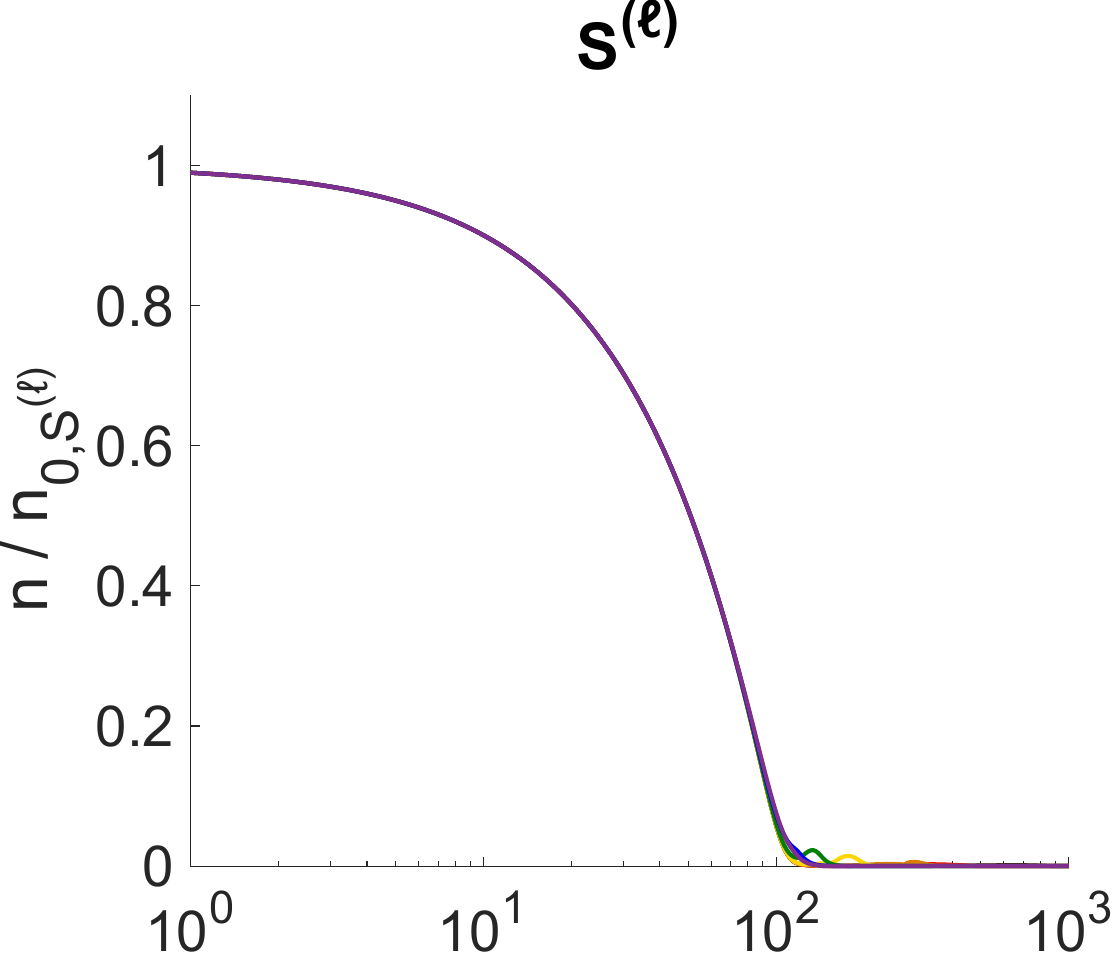}{a}
    \labelpanel[scale=0.35]{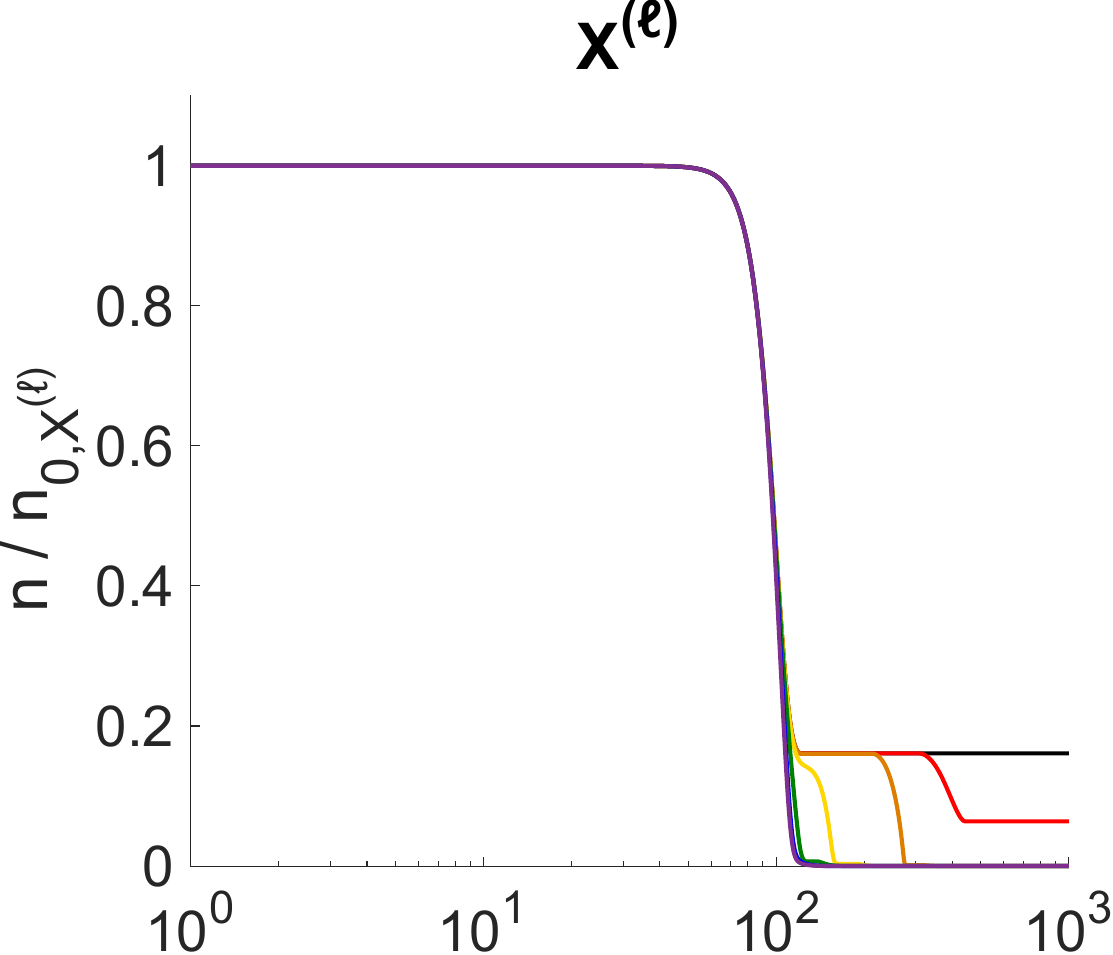}{b}
    \labelpanel[scale=0.35]{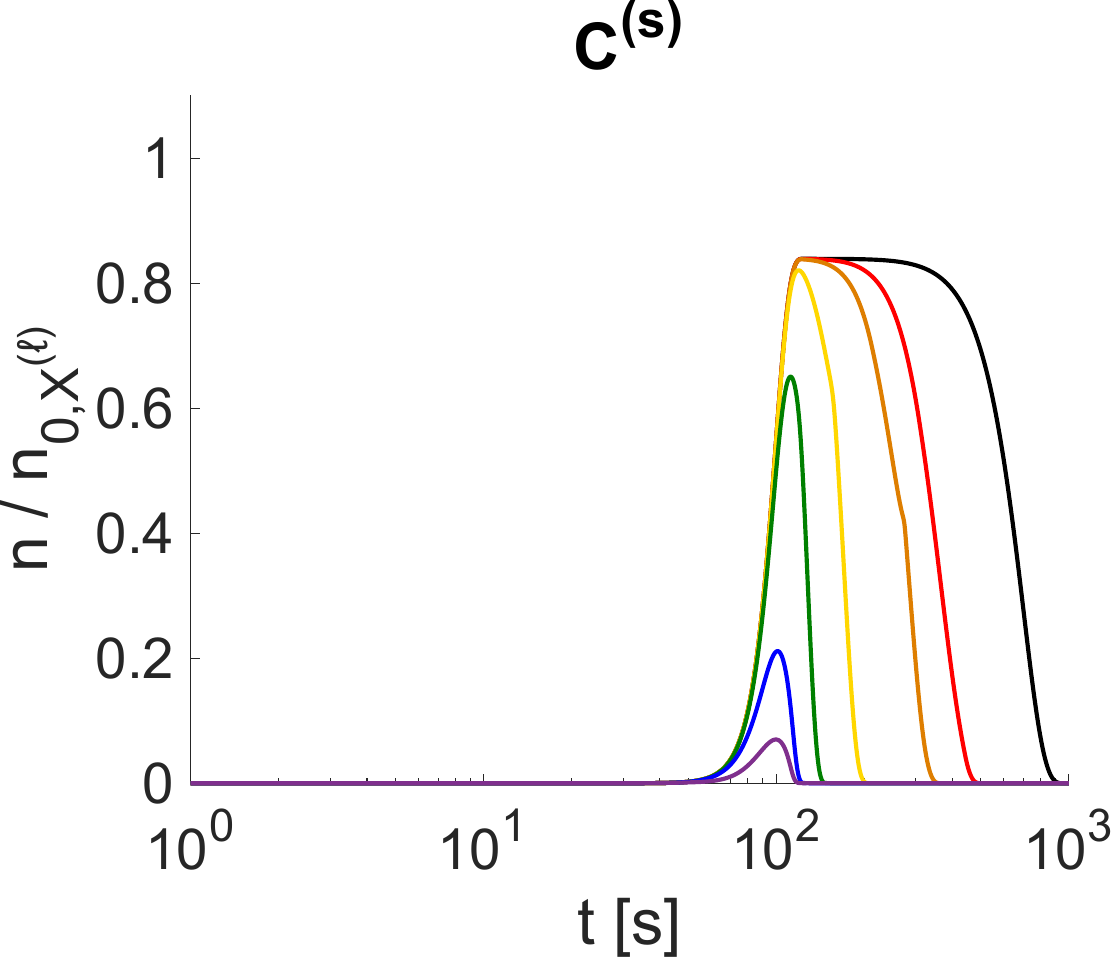}{c}
    \labelpanel[scale=0.35]{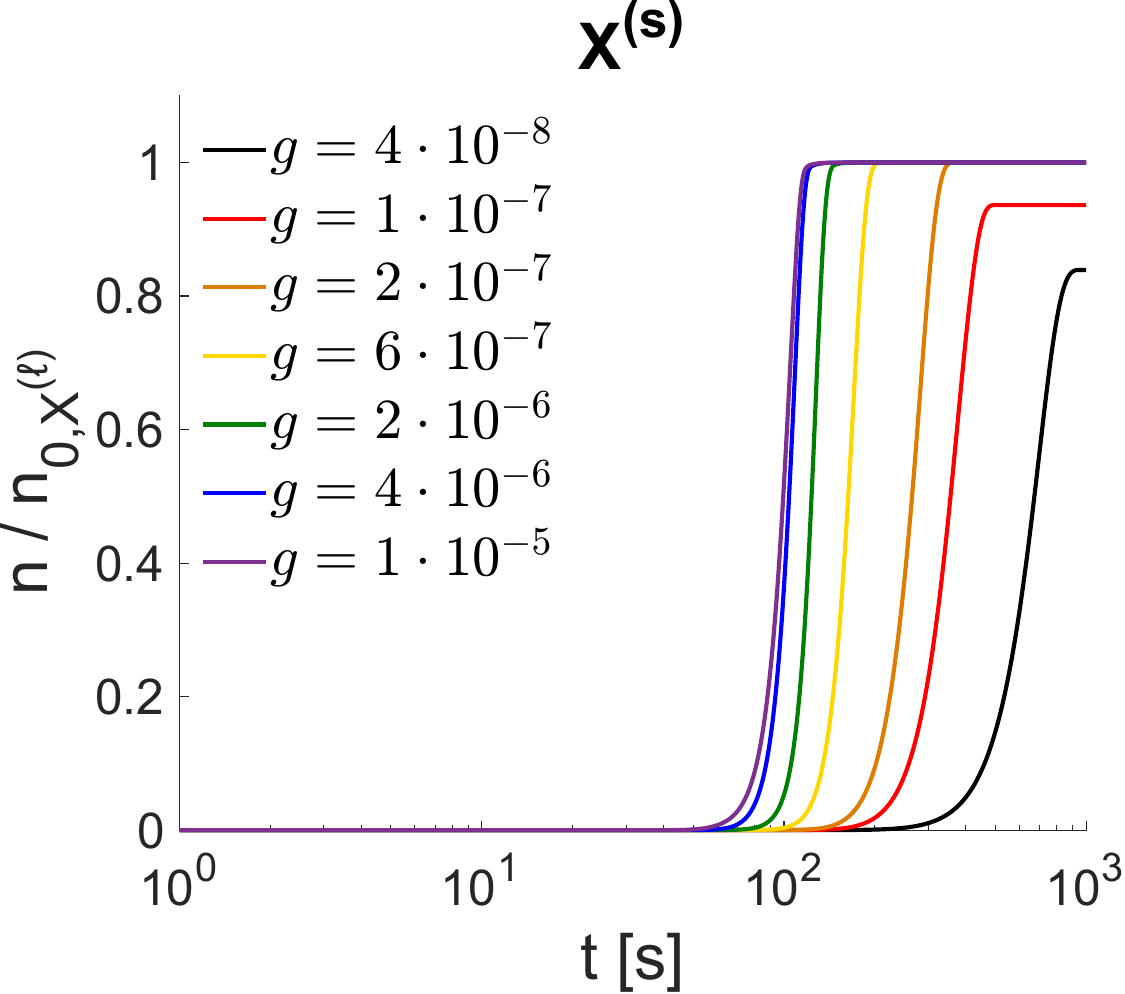}{d}
    \labelpanel[scale=0.35]{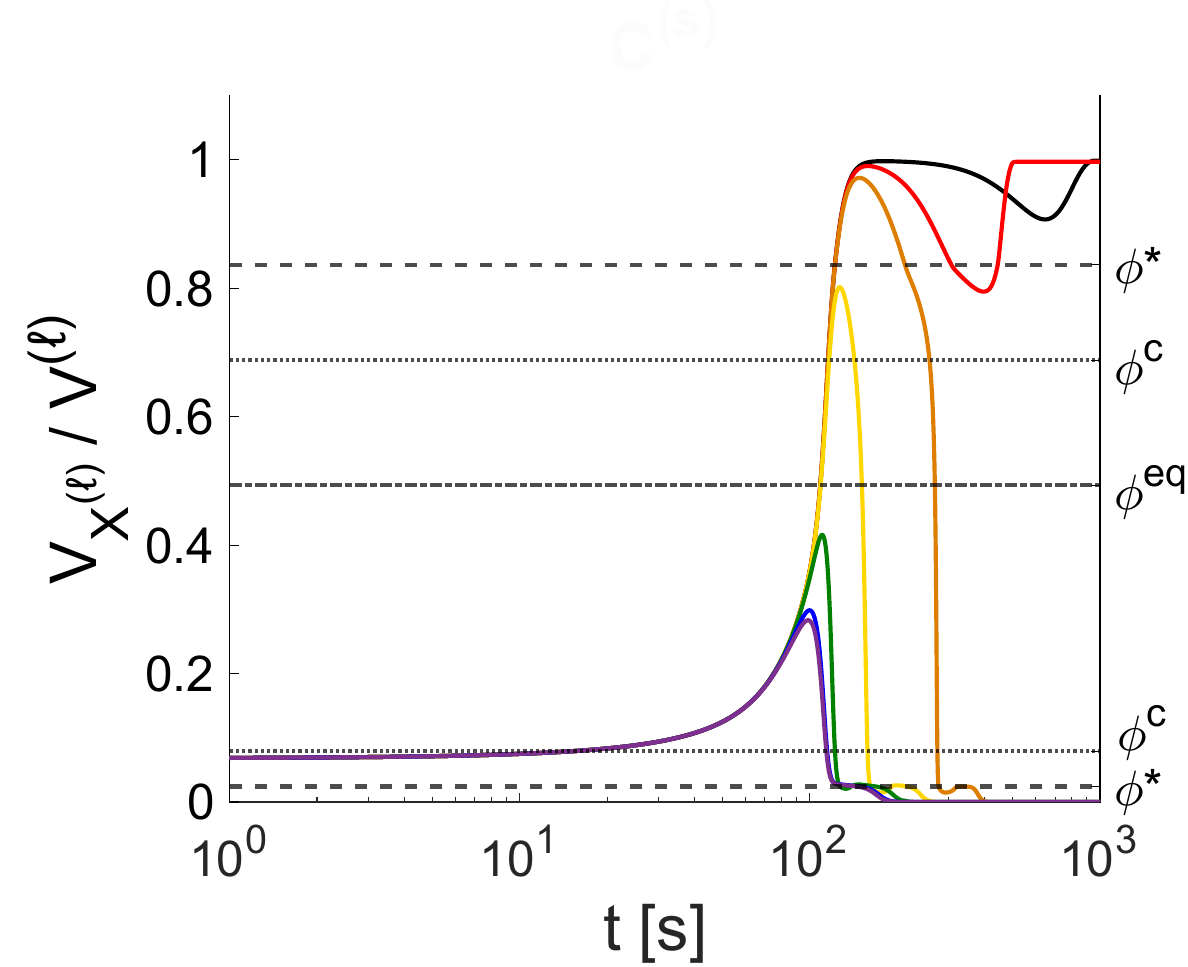}{e}
    \caption{Evaporation-induced successive solvent-coupled crystallization for various growth rates $g$ of the secondary $X^{(s)}$ crystallization and starting from an initially evaporation-induced primary $C^{(s)}$ crystallization that is slower than the evaporation. a,b,c) Relative material-resolved molecule numbers over time. e) Precursor volume fractions in the liquid phase per time. a,b,c) Evaporation of the solvent $S^{(\ell)}$ induces $C^{(s)}$ crystallization, which drains $X^{(\ell)}$ but may not completely consume it (red to black in b). d) $X^{(s)}$ forms from $C^{(s)}$ releasing solvent. If the $X^{(s)}$ crystallization is sufficiently fast (orange to purple), it lifts the solvent limitation from the $C^{(s)}$ crystallization. If it is slow (red to black), however, it can only partially counterbalance solvent evaporation. The $C^{(s)}$ crystallization therefore remains solvent-limited, which results in $X^{(\ell)}$ residues. The LaMer curves in panel e) show two basic behaviors in the significant crystallization regions. For the faster $X^{(s)}$ crystallizations (purple to orange), the concentration of the supersaturated precursor decreases to the saturation limit of the precursor in the solvent, $\phi^* < \phi^{\text{eq}}$, at which the $C^{(s)}$ crystallization subsequently continues. Conversely, for slower $X^{(s)}$ crystallizations (red to black), the solvent $S^{(\ell)}$ is exhausted before the precursor $X^{(\ell)}$. Consequently, the liquid precursor volume-fraction curves approach 1. 
    \label{fig:EvapSucc_Varg2}}
    \end{figure*}

Now, we fix the evaporation kinetics and keep the primary crystallization P2 slower than the evaporation, as for orange and red in \cref{fig:SSPf}.
Subsequently, we vary the speed of the secondary crystallization, P3, via the growth-rate scaling. 
The $X^{(s)}$ curves in \cref{fig:EvapSucc_Varg2} show three regimes. 

If P3 is very slow (black), the recycled $S^{(\ell)}$ evaporates before it can be consumed by P2. P2's kinetics therefore remain solvent-limited: $C^{(s)}$ crystallization stops early, and $X^{(\ell)}$ residues may remain. As a result, $X^{(\ell)}$ to $X^{(s)}$ conversion is incomplete. 

If P3 is slightly faster (red), the solvent limitation on P2 can be partly lifted, resulting in distinct crystallization phases. The LaMer curve in panel b) crosses $\phi^* > \phi^{\text{eq}}$ three times: crystal growth pauses after the first crossing, then resumes between the second and third crossings.
Finally, if P3 is sufficiently fast (orange to purple), solvent recycling due to P3 lifts the solvent limitation on P2. Consequently, $X^{(\ell)}$ is fully converted to $C^{(s)}$ and the $X^{(\ell)}$-to-$X^{(s)}$ conversion is complete. Within this regime, for P3 even faster (purple), P3 is P2-limited, and the $X^{(s)}$ curve follows the $C^{(s)}$ production — as also visible in the corresponding  $X^{(\ell)}$ curve.

\FloatBarrier
\section{Conclusions}
In summary, we developed a general zero-dimensional population balance framework
coupling crystallization, evaporation, and chemical reactions. For perovskite films
formed from solution, we proposed an effective reaction template informed by
multichannel \emph{in situ} studies of MAPI in common solvents, intended as a
reasonable mechanistic prior for these systems. Within that framework, we used
parameter sweeps to investigate the evaporation-driven two-step (successive)
crystallization subsystem, building up from subsystems with fewer coupled processes.

The resulting quantitative evaporation, phase-resolved crystallization, and LaMer
curves yield several findings and their consequences:

i)~Allowing finite-size nuclei, crystallization admits a nucleation-dominated regime
alongside the classical growth-dominated one, lowering the Avrami exponent without
changing the growth dimension. The Avrami exponent is therefore ambiguous as a growth
dimension.

ii)~When a secondary step is kinetically limited by the primary one, the intermediate
phase can be practically invisible, although the mechanism is indirect. The absence of
an intermediate signal is insufficient evidence for direct formation.

iii)~With limited solvent, the primary and secondary solvent-coupled steps are mutually
coupled regardless of their relative rates. In the evaporation-driven case, intermediate
crystallization is therefore either incomplete, from solvent depletion when evaporation
is fast, or rate-limited by evaporation itself rather than by its own kinetics when it
is slow; perovskite can also form at the evaporation rate with no intermediate
detectable at all. Evaporation therefore shapes both phase-specific crystallization
curves --- directly the intermediate's, and, through it, the perovskite's --- so neither
curve identifies the mechanism on its own, and evaporation must be measured with them.

These findings concern the subsystem studied; because they depend mainly on relative
timescales, they should hold irrespective of the rate parameters, and hence of material
system and processing conditions. Varying evaporation rates rather than crystallization
rates, for instance, should give equivalent
results~\ccite{majewskiSimulationPerovskiteThin2024}.

Where diffusion limitations or preferential crystallization at the substrate or film
surface dominate, spatially resolved modeling will be required. Because the template
retains the direct and successive channels in parallel, their competition can be
examined without further model construction; and because mechanistic specificity enters
only through the rate parameters, the same framework applies to other materials and to
the effects of temperature and precursor composition. Ultimately, these mechanistic
signatures offer a route to distinguishing crystallization pathways from the features of
measurable drying and phase-resolved composition curves.

\section*{Acknowledgements}
We acknowledge the financial support from the Helmholtz Association through the Solar Technology Acceleration Platform for emerging Photovoltaics ( SolarTAP ) and the Deutsche Forschungsgemeinschaft (DFG) via the Collaborative Research Center (‘ChemPrint’, Project No. 538767711, CRC 1719).

\section*{Appendix}
\subsection{Model Description}
\begin{figure}[htbp]
    \centering
    \includegraphics[scale=0.35,trim=0 9cm 0 0]{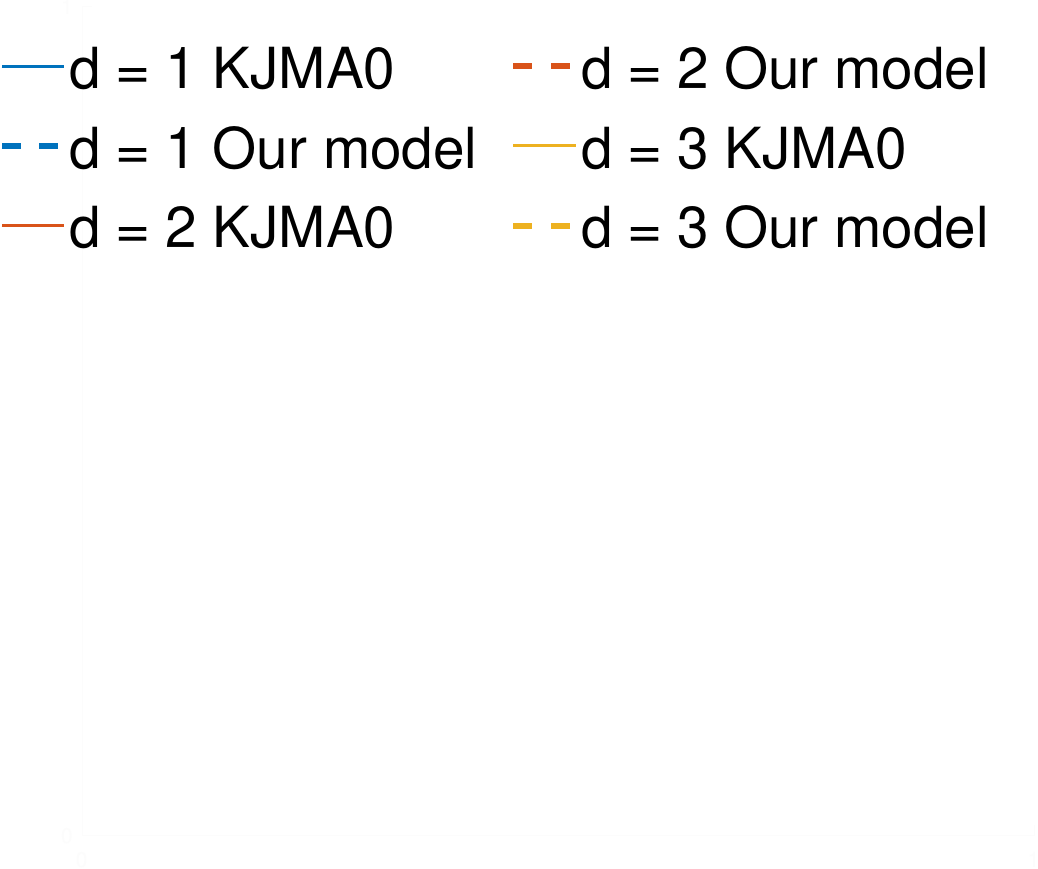}
    \includegraphics[scale=0.35]{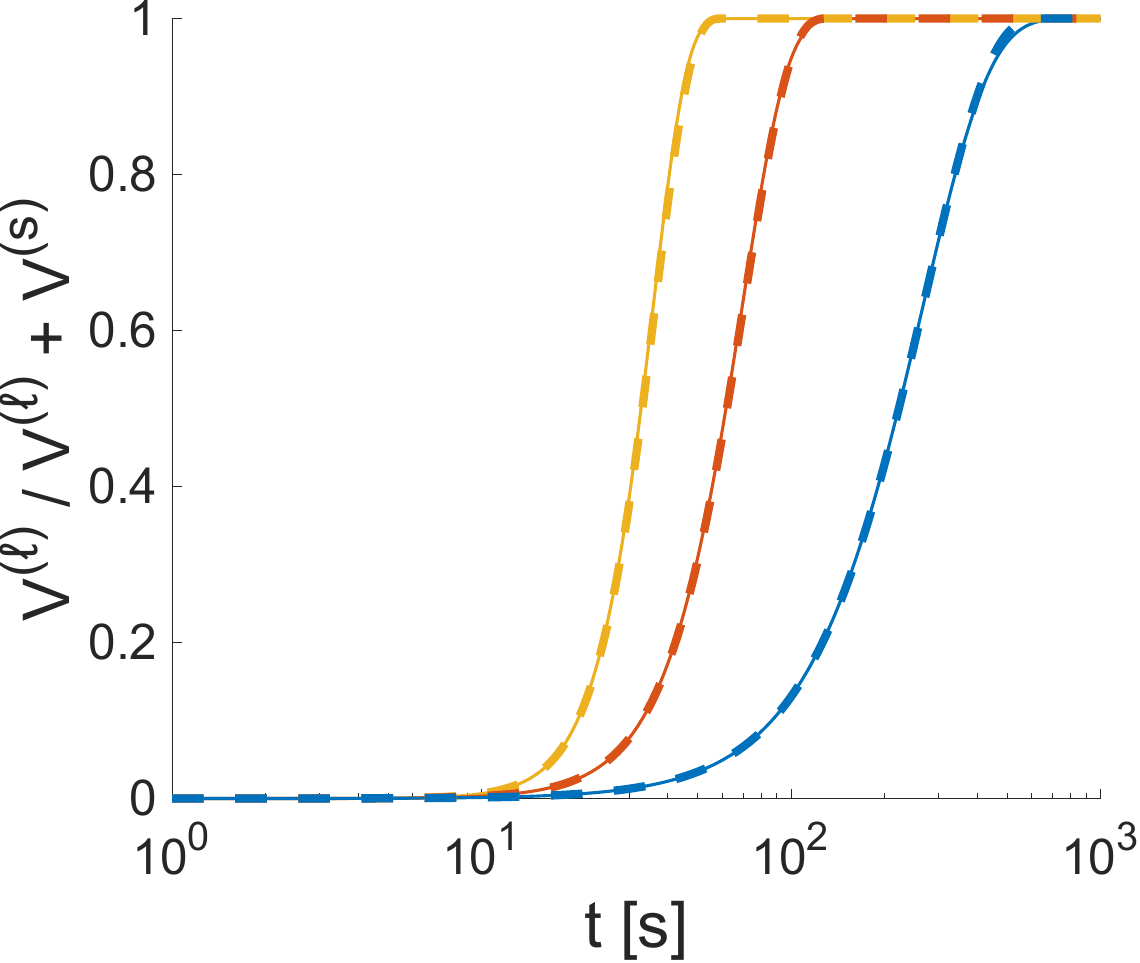} 
    \caption{ Transformed crystalline volume fraction per time for a single crystallization $B^{(\ell)} \to B^{(s)}$ in a $B^{(\ell)}, B^{(s)}$- blend, calculated from our crystallization cell-model and the common literature model KJMA0: $\phi^{(s)}(t) = 1-e^{- B_d \dot{\eta} g^d  t^{d+1}}$.  For a critical radius $r_c =0$, nucleation rate density $\dot{\eta}_0 = 3\cdot 10^{-4}$, and growth rate $ g_0 = 0.2$.   The curves generated with our model using an impingement function $f = (\frac{V^{(\ell)}}{V^{(\ell)} +V^{(s)}})^{0.7}$ closely match the curves from the common KJMA0 model. } 
    \label{fig:impingement function justification}
\end{figure}

\subsection{Model Behaviour \label{app:parTables}}

\begin{table*}[t]
\begin{tabular}{l|llll}
                              & DMF   & PbI2                 & MAI                 & MAPI      \\
\hline 
chemical formula             & $\text{C}_3\text{H}_7\text{NO} $     & $\text{PbI}_2$              &  $\text{CH}_3\text{NH}_4$I  &  $[\text{CH}_3\text{NH}_3]^+ \text{Pb}_2^+{\text{I}}_3^- $         \\
density $\rho$ $[kg/m^3]$     &  944                 &    6160              & 2280                     &  4118        \\
molar mass $M$ $[kg/mol]$      & 0.0731           &   0.4610             & 0.1590                  & 0.6200        \\
vapor pressure $P^*$ $ [Pa]$                & 490.9      & & &  \\
\end{tabular}
\caption{Material characteristics  at $T_0 = 298 \, [K] , P_0 = 10^5 \,  [Pa]$ }
\label{tab:MaterialParameters}
\end{table*}

\paragraph{Simulation Setup}
Each run is specified by the material properties \cref{tab:MaterialParameters}, its initial and boundary conditions, the temperature $T$, Flory parameter $\chi$ and evaporation
prefactor $\alpha$ (Table~\cref{tab:SI_Conditions}), and the parameters of each
crystallization, $C_1$ and, where present, $C_2$ (Table~\cref{tab:SI_Cryst}) according to the rate laws:
\begin{align*}
    \dot{\eta} &=    \dot{\eta}_0'  \,\max \left(\frac{\phi_R - \phi_R^{c}}{1- \phi_R^{c}} ,0 \right),   \\
    g &=    g_0' \, \,\max \left(\frac{ \phi_R - \phi_R^{*} }{1-\phi_R^{*} },0 \right).
\end{align*}
For $\phi_R^c<1$ and $\phi_R^*<1$, and using $\dot\eta_0'=\dot\eta_0(1-\phi_R^c)$ and $g_0'=g_0(1-\phi_R^*)$, these are identical to \cref{eq:rateParameterizations}. In the main text, the factors $(1-\phi_R^c)^{-1}$ and $(1-\phi_R^*)^{-1}$ have been absorbed into the scalings $\dot\eta_0$ and $g_0$, respectively, for ease of presentation. 
The reaction volume fraction $\phi_{R_i}$ entering~\cref{eq:rateParameterizations} is
supplied by one of three closures: $\phi_{R_i}=\phi_1$ for the pure-material runs, $\phi_{R_1}=2.25\,\phi_1\phi_2$ where crystallisation $C_1$ is solvent-coupled, and $\phi_{R_2}=\phi_3$ for $C_2$.  A cell marked \sweep\ indicates a parameter that is varied across the curves of the figure/
panel if specified, with the per-curve values given in the figure legend. Quantities
are dimensionless if no unit is noted.

Parameters were generally chosen so that the effects discussed in the main text are visible within a single observation window.    

\begin{table}
\centering
\footnotesize
\setlength{\tabcolsep}{5pt}\renewcommand{\arraystretch}{1.2}
\begin{tabular}{@{}cl l l c c c@{}}
Fig. &  Processes & $\phi_0$ & $T$/ [K] & $\chi$ & $\alpha$ \\
\midrule
3 &  $\phantom{E_1,}C_1$        & $(1\phantom{.000},0\phantom{.000\,},\,0,\,0)$ &  &  &  \\
4 &  $\phantom{E_1,}C_1$        & $(\sweep\phantom{00},\sweep\phantom{00\,},\,0,\,0)$ &  &  &  \\
5 &  $\phantom{E_1,}C_1,C_2$    & $(1\phantom{.000},0\phantom{.000\,},\,0,\,0)$ &  &  &  \\
6 &  $\phantom{E_1,}C_1,C_2$    & $(\sweep\phantom{00},\sweep\phantom{00\,},\,0,\,0)$ &  &  &  \\
7 &  $C_1,C_2$    & $(0.25\phantom{0}, 0.75\phantom{0\,},\,0,\,0)$ &  &  &  \\
8 & $E_1,C_1$    & $(0.068,\,0.932,\,0,\,0)$ & $351$ & $0$ & $10^{-5.3578}$ \\
9 &  $E_1,C_1,C_2$& $(0.068,\,0.932,\,0,\,0)$ & $352$ & $0$ & $10^{-5.3578}$ \\
10 & $E_1,C_1,C_2$& $(0.068,\,0.932,\,0,\,0)$ & $352$ & $0$ & $10^{-5.3578}$ \\
\end{tabular}
\caption{\textbf{Initial and boundary conditions.} $\phi_0$ is the initial composition with solute, solvent, solid-state intermediate, perovskite components; $\chi$ are the pairwise Flory–Huggins interaction parameters and $\alpha$ is the prefactor appearing in the rate expression~\cref{eq:Evap} for the evaporation $E_1$, absent where no evaporation is
active. }
\label{tab:SI_Conditions}
\end{table}

\begin{table}
\centering
\footnotesize
\setlength{\tabcolsep}{2pt}\renewcommand{\arraystretch}{1.2}
\begin{tabular}{@{}cl l cccc cc@{}}
Fig. & Cryst. & $\phi_{R_i}$ & $\dot{\eta}_0' / [\#/m^d]$ & $g_0' / [m/s]$ & $d$ & $r_c / [m]$
     & $\phi_R^{*}$ & $\phi_R^{c}$ \\
\midrule
\multicolumn{9}{@{}l}{\itshape One crystallisation}\\
\addlinespace[2pt]
3 a) & $C_1$ & $\phi_1$  & $10^{18}$ & \sweep  & $3$ & $10^{-10}$   & $0$ & $0$ \\
3 b) & $C_1$ & $\phi_1$  & \sweep & \sweep  & \sweep & $10^{-10}$   & $0$ & $0$ \\
3 c,d) & $C_1$ & $\phi_1$  & \sweep & \sweep & $3$ & $10^{-10}$   & $0$ & $0$ \\
4 & $C_1$ & $2.25 \, \phi_1\phi_2$  & $10^{13}$ & $10^{-6}$ & $3$ & $10^{-8}$ & $0.1$ & $0.25$ \\
\addlinespace
\multicolumn{9}{@{}l}{\itshape Two crystallisations}\\
\addlinespace[2pt]
5 & $C_1$ & $2.25 \, \phi_1\phi_2$
  & $10^{4}$ & $10^{-8}$ & $1$ & $10^{-8}$ & $0$ & $0$ \\
5 & $C_2$ & $\phi_3$
  & $10^{10}$ & \sweep & $3$ & $10^{-8}$ & $0$ & $0$ \\
6 & $C_1$ & $2.25 \, \phi_1\phi_2$
  & $10^{5}$ & $10^{-6}$ & $1$ & $10^{-8}$ & $0.05$ & $0.15$ \\
6 & $C_2$ & $\phi_3$
  & $10^{12}$ & $3\!\cdot\!10^{-6}$ & $3$ & $10^{-8}$ & $0.1$ & $0.25$ \\
7 & $C_1$ & $2.25 \, \phi_1\phi_2$
  & $10^{4}$ & $10^{-6}$ & $1$ & $10^{-8}$ & $0.1$ & $0.25$ \\
7 & $C_2$ & $\phi_3$
  & $10^{12}$ & \sweep & $3$ & $10^{-8}$ & $0.1$ & $0.25$ \\
\addlinespace
\multicolumn{9}{@{}l}{\itshape Evaporation active ($E_1$, Table~\ref{tab:SI_Conditions})}\\
\addlinespace[2pt]
8 & $C_1$ & $2.25 \, \phi_1\phi_2$
  & $10^{5}$ & \sweep & $3$ & $10^{-8}$ & $0.05$ & $0.15$ \\
9 & $C_1$ & $2.25 \, \phi_1\phi_2$
  & $10^{11}$ & $5\!\cdot\!10^{-5}$ & $3$ & $10^{-8}$ & $0.05$ & $0.15$ \\
9 & $C_2$ & $\phi_3$
  & $10^{11}$ & \sweep & $3$ & $10^{-8}$ & $0.05$ & $0.15$ \\
10 & $C_1$ & $2.25 \, \phi_1\phi_2$
  & $10^{11}$ & $4\!\cdot\!10^{-6}$ & $3$ & $10^{-8}$ & $0.05$ & $0.15$ \\
10 & $C_2$ & $\phi_3$
  & $10^{11}$ & \sweep & $3$ & $10^{-8}$ & $0.05$ & $0.15$ \\
\bottomrule
\end{tabular}
\caption{\textbf{Crystallisation parameters} $\dot\eta_{0}' = \dot\eta_{0} \cdot (1-\phi_R^c)$, $g_{0}' = g_0 \cdot (1-\phi_R^*)$ are the resealed nucleation rate density and growth rate scalings, $d$ is growth dimension, $r_{c}$ the critical radius and
$\phi_{R}^{*}$, $\phi_{R}^{c}$ the growth and nucleation thresholds.
$\phi_{R_i}$ is the reactant volume fraction given in the third column.}
\label{tab:SI_Cryst}
\end{table}

\paragraph{Numeric}
All evolution equations were solved using a first-order Euler discretization with rate-clipping safeguards to ensure positivity while preserving population number conservation. Domain constraints for physical quantities have generally been enforced explicitly. Simulation results have been cross-checked for several time resolutions. Physical consistency against the expectations based on the reaction systems has been assessed in all cases. Where possible, implementations have been validated against analytical results, e.g., \cref{fig:impingement function justification}.

\clearpage
\bibliographystyle{custom-apsrev4-2}
\bibliography{Bibliography}
\end{document}